\documentclass[12pt]{article}
\usepackage[dvips]{graphicx}
\usepackage{latexsym}
\usepackage{pdproc}

  \makeatletter 
  \def\@cite#1{[#1]} 
  \makeatother    
\def\err#1#2{\lower2pt\hbox{ $\stackrel{\scriptstyle +#1}{\scriptstyle
-#2}$}}
\newcommand{\agt}{\mathrel{\raisebox{-.6ex}{$\stackrel{\textstyle>}{\sim}$}}}
\begin{document}
\renewcommand{\thefootnote}{\alph{footnote}}

\title{Collider Phenomenology for a few models of extra dimensions
\footnote{
Plenary talk given at the 12th International Conference
on Supersymmetry and Unification of Fundamental Interactions (SUSY 2004)
June 17-23, 2004, Epochal Tsukuba, Tsukuba, Japan.
}
}
\author{KINGMAN CHEUNG}
\address{Department of Physics and NCTS, National Tsing Hua
University, Hsinchu, Taiwan, R.O.C. \\
{\rm E-mail: cheung@phys.nthu.edu.tw}
}

\abstract{
In this talk, we summarize the collider phenomenology and recent experimental
results for various models of extra dimensions, including the large
extra dimensions (ADD model), warped extra dimensions (Randall-Sundrum model),
TeV$^{-1}$-sized extra dimensions with gauge bosons in the bulk,
universal extra dimensions, and an 5D SU(5) SUSY GUT model in AdS space.
}

\normalsize\baselineskip=15pt

\section{Introduction}

The standard model (SM) of particle physics can be considered the
most successful model among  various standard models (other standard
models, e.g., the standard model of the sun and the standard model
of cosmology are gaining ground as more and more data are
available to refine the models. From now on the standard model is
referred to the standard model of particle physics.)  It has
enjoyed great health for more than 30 years.  The precision
measurements at LEP have tested the SM to the level of $10^{-3}$ \cite{lep}.
In addition, the last piece of quarks, the top quark, was found \cite{top}.
Nevertheless, as a theorist we believe the SM cannot be a final
theory, because of the followings. (i) The SM has many parameters,
most of which are the fermion masses.  This is related to the
flavor problem.  In the SM, we have three generations of fermions,
each of which seems to be a repetition of each other.  We do not
fully understand why it is so and why there are only three
generations, not to mention the generation of the fermion mass
pattern.  (ii) The SM is not a real unification of all forces.  It
would be nice to embed the SM into a grand unification theory.
(iii) The hierarchy problem tells us that the apparently only two
scales in particle physics, the electroweak and Planck scales, are
$16-17$ orders of magnitude different, which gives an enormously
large loop correction to the scalar boson mass.  It requires
a very precisely fine-tuned bare mass to cancel the loop
correction in order to give a scalar boson mass of order $O(100)$
GeV.  All these problems lead us to believe that there should be
new physics beyond the SM.  Most of us believe that the new
physics should come in in the TeV scale.  There is a hope that the
upcoming LHC is the place for the next big discovery.

There are other observations that tell us that the SM is not
satisfactory.  The most striking evidence is the definite (though
small) neutrino masses that are required in the neutrino
oscillations.  There are mounting evidences for the solar and
atmospheric neutrino flux deficits that are best explained by
neutrino oscillations.  Most of us also believe that there should
be dark matter that fills up a substantial fraction of the
universe.  More surprisingly, there is another very mysterious
component of the universe, which is revealed by recent
balloon, supernova, and satellite experiments.  It is clear 
that the SM cannot provide these components of the universe.  In
addition, the SM cannot fulfill all the requirements to give a
sufficiently large enough baryon asymmetry of the universe.

\begin{figure}[ht!]
\centering
\includegraphics[width=5in]{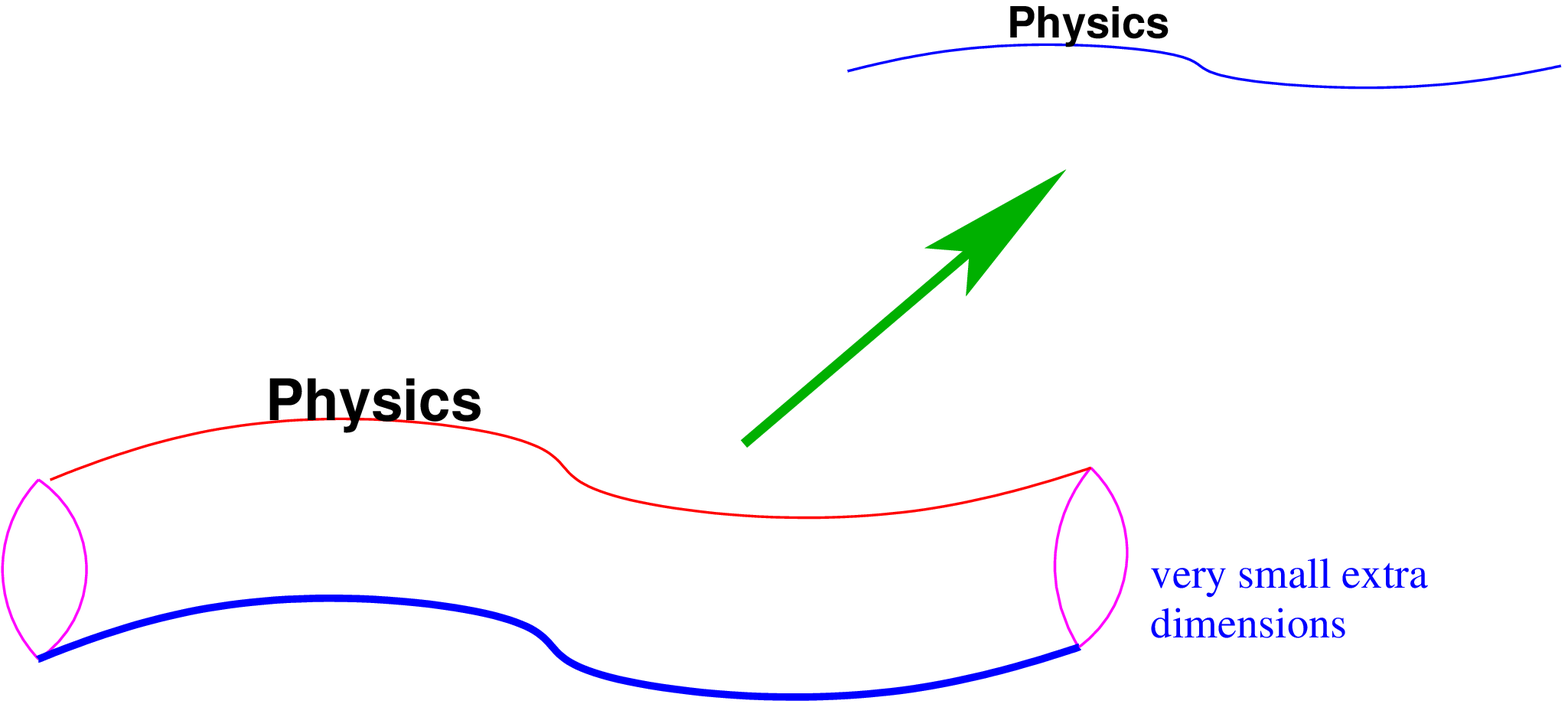}
\caption{\label{cylinder}
Figure illustrating why we do not see the extra dimensions.}
\end{figure}

The hierarchy problem has motivated a number of models beyond the
SM.  In recent years, a number of models in extra dimensions have
been proposed.  They provide an alternative view of the hierarchy
problem into a geometric stabilization of the extra dimensions.
If there exist
extra dimensions, why do we not see them?  One simple reason is
that they are probably too small.  Figure \ref{cylinder}
illustrating this simple reason why we do not see the extra
dimensions.  The word ``Physics" is sitting on the cylinder (extra
dimensions), but when we see it from very far away, the cylinder
is too small to be noticed.

The main purpose of this talk is to review the collider
phenomenology associated with extra dimension models, both
theoretical and experimental works.  Subsequent sections are
devoted to various models, namely, (i) the large extra dimension
model (ADD model), (ii) the warped extra dimension model
(Randall-Sundrum mode), (iii) TeV$^{-1}$-sized model with gauge
bosons in the bulk, (iv) universal extra dimension model, and (v)
an 5D SU(5) SUSY GUT model on a slice of AdS space.

\section{ADD model}

\begin{figure}[ht!]
\centering
\includegraphics[width=4.5in]{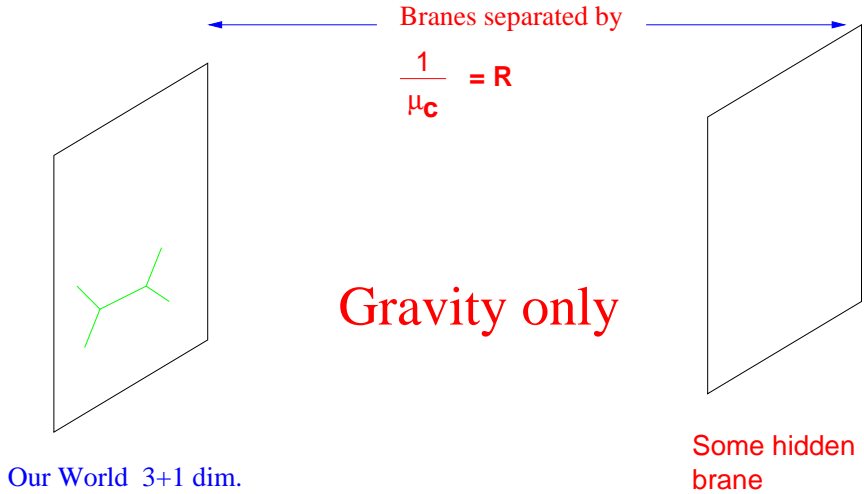}
\caption{\small The setup in the ADD model \label{add}}
\end{figure}

\begin{figure}[ht!]
\centering
\includegraphics[width=4.5in]{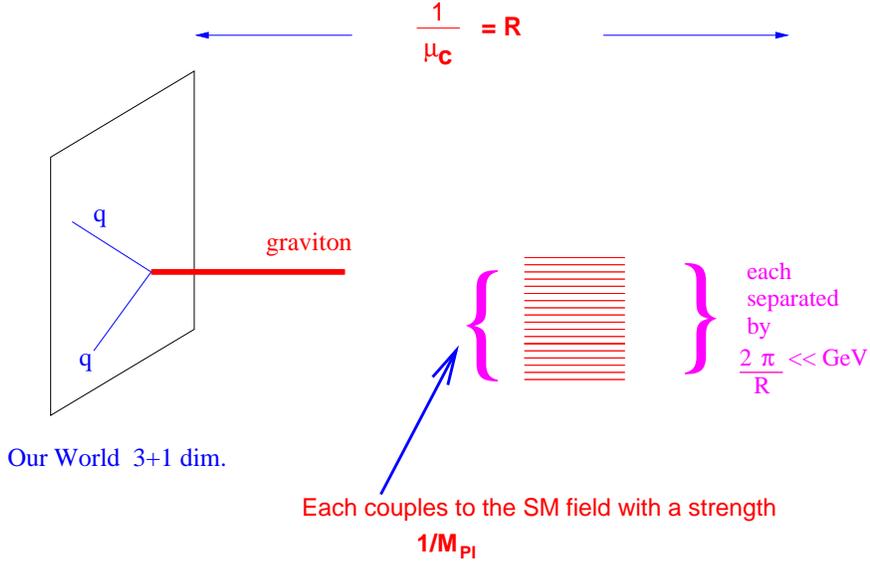}
\caption{\small The interactions of the graviton in the ADD model
\label{add-2}}
\end{figure}

It was proposed by Arkani, Dimopoulos, and Dvali \cite{arkani}
that the size $R$
of the extra dimensions that only gravity can propagate can be as
large as mm.  This observation was based on the fact that no
deviation from the Newton's law has been observed down to mm size.
It has an important impact in our understanding of gravity.
Suppose the fundamental Planck scale of the model is $M_D$, the
observed Planck scale $M_{\rm Pl}$ becomes a derived quantity:
\[
M^2_{\rm Pl} \sim M_D^{n+2} \; R^n \;,
\]
where $R$ is the size of the extra dimensions.  This expression
tells us that if $R$ is extremely large, as large as a mm, the
fundamental Planck scale $M_D$ can be as low as TeV.  Since the
fundamental Planck scale is now at TeV, the hierarchy problem no
longer exists.  The setup is shown in Fig. \ref{add}.  In this
model, the SM particles and fields are confined to a brane while
only gravity is allowed to propagate in the extra dimensions.
Thus, the only probe of the extra dimensions must be through the
graviton interactions, which is illustrated in Fig. \ref{add-2}.
The graviton in the extra dimensions is equivalent to a tower of
Kaluza-Klein (KK) states in the 4D point of view with a mass
spectrum given by
\[
M_l = \frac{l}{R} \;,
\]
where $l=0,1,2,...$.  The separation between each state is of
order $1/R$, which is very small of order of $O(10^{-4})$ eV. This
means that in the energy scale of current high energy experiments,
the mass spectrum of the KK tower behaves like a continuous
spectrum.  Each of the KK states interacts with a strength of
$1/M_{\rm Pl}$ with the SM particles.  However, when all the KK
states are summed up, the interaction has a strength of $1/M_D$,
where $M_D$ is the real fundamental scale of the model and of
order of $O(TeV)$.  Thus, we may be able to detect the graviton
effects in current and future high energy experiments.

The collider signatures for the ADD model can be divided into (i)
sub-Planckian and (ii) trans-Planckian. {}From Fig.
\ref{add-2}, it is clear that only graviton can probe the extra
dimensions when the energy scale is below $M_D$.  The SM particles
scatter into a graviton, which can either (i) go into the extra
dimensions and does not come back to the brane, which then gives
rise to missing energy and momentum in experiments, or (ii) come
back to the SM brane and decay back into SM particles, the
scattering amplitude of which then interferes with the normal SM
amplitude.  Therefore, experimentally we can search for two types
of signatures, the missing energy or the interference effects.
When the energy scale is above the $M_D$, we expect the quantum
gravity effects become important and objects, like black
holes, string balls, $p$-branes would appear.  In the following, we
shall discuss these signatures one by one.

\subsection{Sub-Planckian}

There have been enormous large amount of literature in this area 
\cite{add-1,add-2,add-3,add-4,add-5,add-6,add-7,add-8},
so I only highlight on some of them and certainly show personal
preference. As mentioned above the sub-Planckian signatures can be
further categorized into those involving graviton emission and
the virtual graviton exchanges.  

{\it Graviton emission}.  Let us first discuss the
processes with graviton emission. The cleanest and easiest-to-see
signature would be a single photon or a gauge boson with the
missing energy due to the disappearance of the graviton into the
extra dimensions \cite{add-3,add-7,add-8}. The processes are
\[
e^+ e^- \to \gamma (Z) G
\]
at $e^+ e^-$ colliders and
\[
q \bar q (g g) \to g G
\]
at hadronic colliders.

\begin{figure}[ht!]
\centering
\includegraphics[width=4.5in]{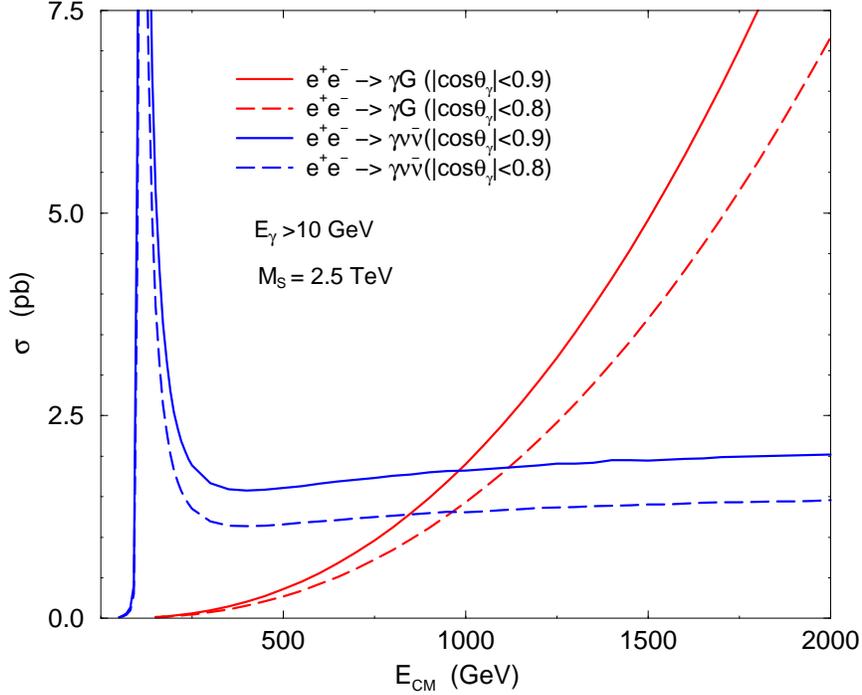}
\caption{Cross sections for $e^+ e^- \to \gamma G$ compared
with the SM background $e^+ e^- \to \gamma \nu \bar \nu$.  From
Ref.~\protect\cite{add-7}.
\label{eeG}}
\end{figure}

In Fig. \ref{eeG}, we show the production cross section for $e^+
e^- \to \gamma G$ vs the center-of-mass energy and compare with
the SM background of $e^+ e^- \to \gamma \nu \bar \nu$.  The graviton signal
easily surpasses the background at $\sqrt{s} \sim 0.8-1$ TeV for a
$M_S =2.5$ TeV. \footnote{There are a few conventions of the
fundamental scale in literature.  They are related to each other
through multiplicative constants. e.g.,
$M_D = [ (2\pi)^n/(8\pi) ]^\frac{1}{n+2}\, M_S$.
}  Other
related processes include $e^+ e^- \to Z G$ and $e^+ e^- \to f
\bar f G$.  

The LEP collaborations 
have searched for $e^+ e^- \to \gamma G^{(k)}$
at $\sqrt{s}=181-208$ GeV.  The L3 Collaboration has the best limit because
of its photon capability \cite{ask}.  The L3 limits are 
$M_D > 1.5 - 0.51$ TeV for $n=2-8$ \cite{ask}. The combined
LEP limits are $M_D>1.6$ TeV for $n=2$ and $M_D>0.66$ TeV for $n=6$ \cite{ask}.
The CDF collaboration also searched for events with a single photon plus
missing energies, but no deviation is observed and they put a
limit on $M_S > 0.55 - 0.6$ TeV for $n=4-8$ \cite{add-cdf}.  
The D\O\ collaboration
searched for events of a single jet with missing energies.  The
limit they obtained is $1-0.6$ TeV for $n=2-7$ \cite {add-d0}.

{\it Virtual graviton exchange.}
For the processes involving virtual graviton exchanges the
signatures would be the interference with the SM amplitudes,
resulting in enhancement in the cross sections, especially at high
energy \cite{add-1,add-2}.  The signals that are easiest to detect are hadronic
dilepton \cite{add-4,add-10,add-10-3} and diphoton production 
\cite{add-11,add-11-1,add-10-3}, 
fermion-pair production at $e^+ e^-$ colliders 
\cite{add-10,add-10-1,add-10-2,add-10-4}.
Other avenues include
gauge-boson pair production \cite{add-12,add-13,add-13-1}, 
dijet production \cite{add-15}, and top-quark production \cite{add-16}, 
as well as the anomalous  muon magnetic moment \cite{add-14}.

\begin{figure}[th!]
\centering
\includegraphics[width=4.5in]{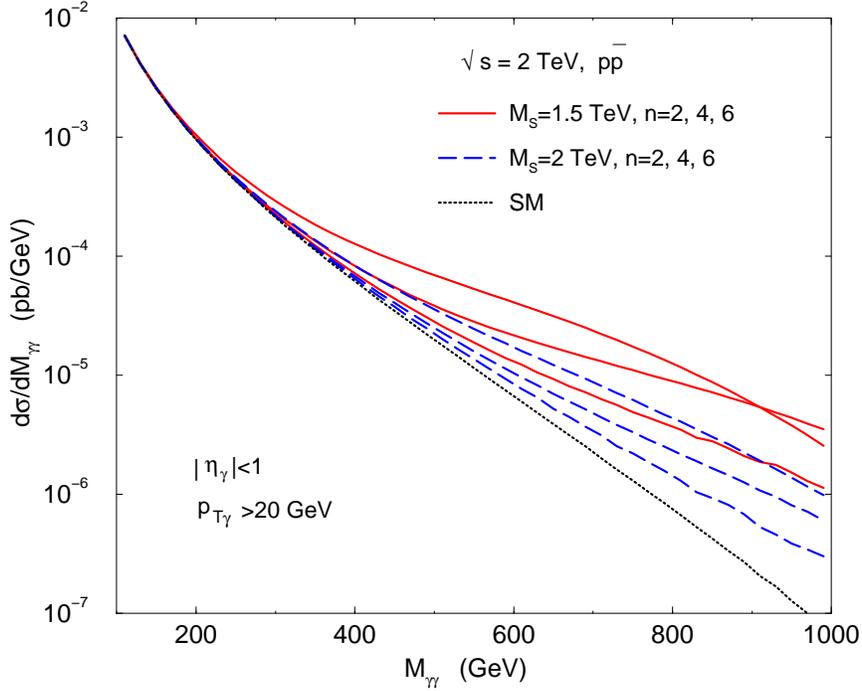}
\caption{\small The diphoton invariant mass spectrum for the
process $p \bar p \to \gamma \gamma$ at the 2 TeV Tevatron.  From 
Ref.~\protect \cite{add-11}.
\label{diphoton} }
\end{figure}

  Figure
\ref{diphoton} illustrates the effect of graviton exchanges in the
diphoton production at the Tevatron.  Enhancement of the cross
section can be seen at $M_{\gamma\gamma}$ much smaller than $M_S$.
Therefore, the process can probe a fundamental Planck scale
considerably higher than the energy of the collider.  Another
interesting process is the light-by-light scattering \cite{add-11}.  The SM
amplitude has to go through a box diagram while the
graviton-induced amplitude is at tree-level.  Thus, the effect of
graviton exchanges is more conspicuous.  Other processes that have
been searched in experiments include $e^+ e^- \to \gamma\gamma,
WW, ZZ$ at LEPII (see the review in Ref. \cite{review3}), 
DIS scattering at HERA, as well as
diphoton$+$dilepton production at the Tevatron by D\O.  We are
going to give more details on the last one, because it gives the
best limit on $M_S$ so far.

\begin{figure}[th!]
\centering
\includegraphics[width=4.5in]{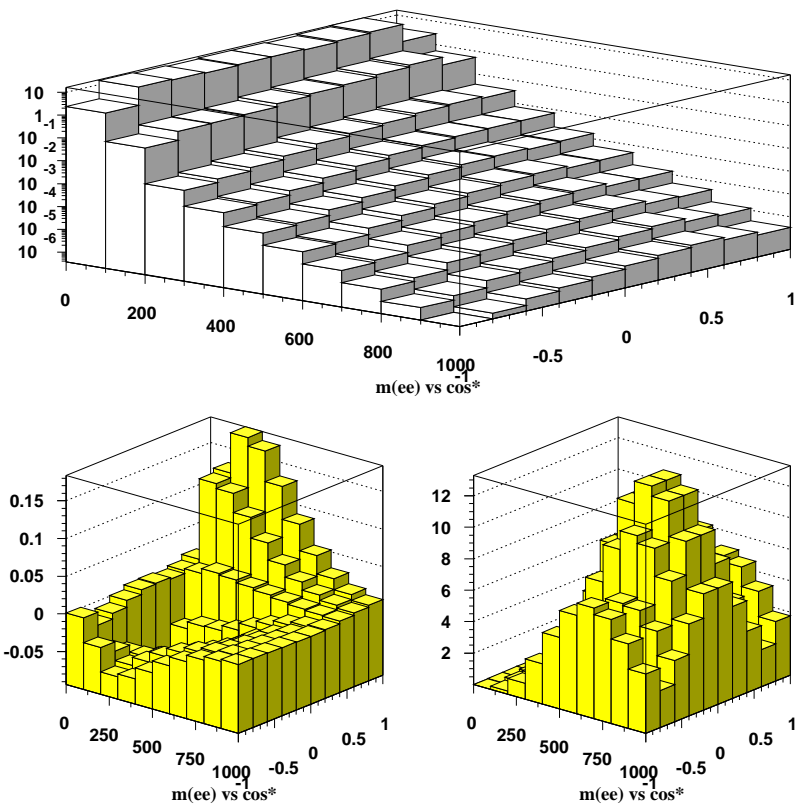}
\caption{\small The dilepton invariant mass spectrum for the
process $p \bar p \to \ell^+ \ell^-$ at the 2 TeV Tevatron.  From 
Ref.~\protect\cite{add-10-3}.
\label{dilepton} }
\end{figure}

\begin{figure}[th!]
\centering
\includegraphics[width=4.5in]{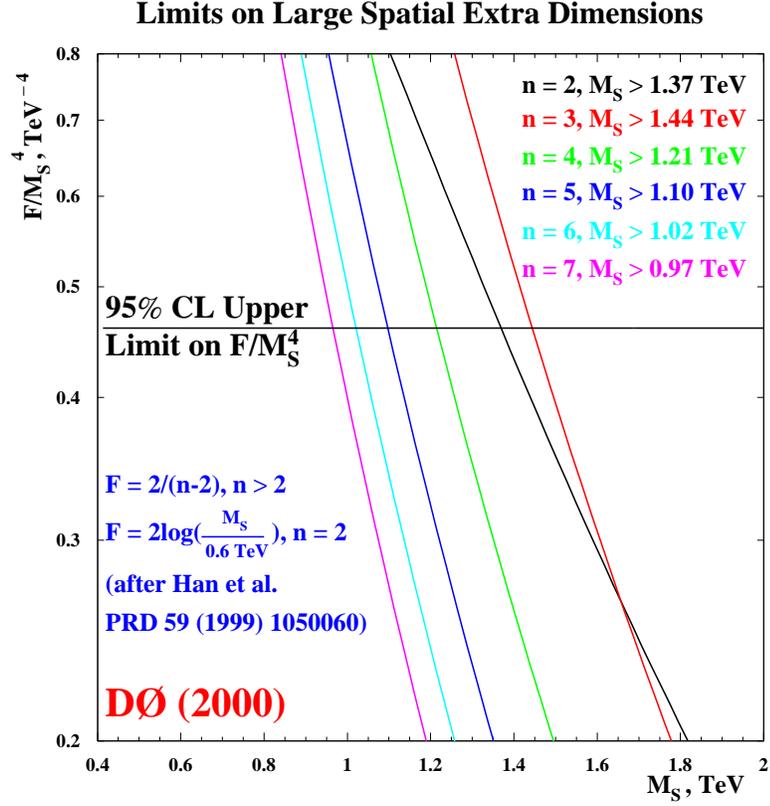}
\caption{\small The limits on $M_S$ obtained by the D\O\
collaboration. This is taken from Ref.~\protect\cite{d0-di}.\label{d0-limit}}
\end{figure}

Cheung and Landsberg \cite{add-10-3} improved the previous analysis on diphoton
and dilepton production using the 2D spectrum,
$d^2\sigma/d\cos\theta^* dM$, where $\theta^*$ is the
center-of-mass frame scattering angle and $M$ is the invariant
mass of the photon or lepton pair.  The advantage of using a 2D
spectrum is that for a $2\to2$ process two kinematical variables
can cover all the phase space, therefore there is no need for cuts
to optimize the effect.  In Fig. \ref{dilepton}, we show the 2D
spectrum of the dilepton process.  It is clear that the
interference term and the pure graviton term are very different
from the SM term.  Moreover, a photon and an electron behave very
similarly to each other in the detector.  Therefore, instead of
losing efficiency in identifying them, we can simply take both of
them  as events of electromagnetic showers. 
D\O\ \cite{d0-di} used this
approach to search for the signal, but the data agreed well the SM
and they placed limits on $M_S$.  The limits are shown in Fig.
\ref{d0-limit}.  The published limit is from $0.97$ to $1.44$ TeV for
$n=2-7$.  
There are some updates in the summer of 2004 from the ICHEP 2004 conference.
The D\O\ preliminary result by combining the Run I and Run II data 
improves the limit to $M_S > 1.70 - 1.14$ TeV for $n=2-7$ \cite{kajf}.  
So far, this is the best limit. 
CDF also updated their limit to $M_S >1.17- 0.79$ TeV for $n=3-7$ \cite{kajf}.
There are also some experimental limits from HERA \cite{kajf}, but are not
as good as the limits from D\O .
We can also study the
sensitivity reach on $M_S$ in the future collider experiments at
Tevatron Run II and at the LHC \cite{add-10-3}, which are shown in Table
\ref{table1}.

\begin{table}[th!]
\centering
\caption{\label{table1} Sensitivity reach on $M_S$ at
Tevatron Run II and at the LHC. }
 \medskip
 \begin{tabular}{|l|cccccc|}
\hline
& \multicolumn{6}{c|}{$M_S\;\;$ (TeV)}\\
&  $n=2$ & $n=4$ & $n=6$  &  $n=2$ & $n=4$ & $n=6$  \\
\hline \hline
&\multicolumn{3}{c}{\underline{Run I}}& & & \\
Dilepton & 1.2 &  1.1 &  0.93  &&&\\
Diphoton & 1.4 &  1.2 &  1.0   &&&\\
Combined & 1.5 &  1.3 &  1.1   &&&\\
\hline &\multicolumn{3}{c}{\underline{Run IIa}} &
                  \multicolumn{3}{c|}{\underline{Run IIb}}  \\
Dilepton & 1.9 & 1.6 & 1.3  & 2.7 & 2.1 & 1.8  \\
Diphoton & 2.4 & 1.9 & 1.6  & 3.4 & 2.5 & 2.1  \\
Combined & 2.5 & 1.9 &
1.6 &3.5 & 2.6 & 2.2 \\
\hline
& \multicolumn{3}{c}{\underline{LHC}} & & & \\
Dilepton & 10 &  8.2 &  6.9  &&&\\
Diphoton & 12 &  9.5 &  8.0  &&&\\
Combined &13  & 9.9 &
8.3 &&&\\
\hline
\end{tabular}
\end{table}

\begin{figure}[th!]
\centering
\includegraphics[width=3.3in]{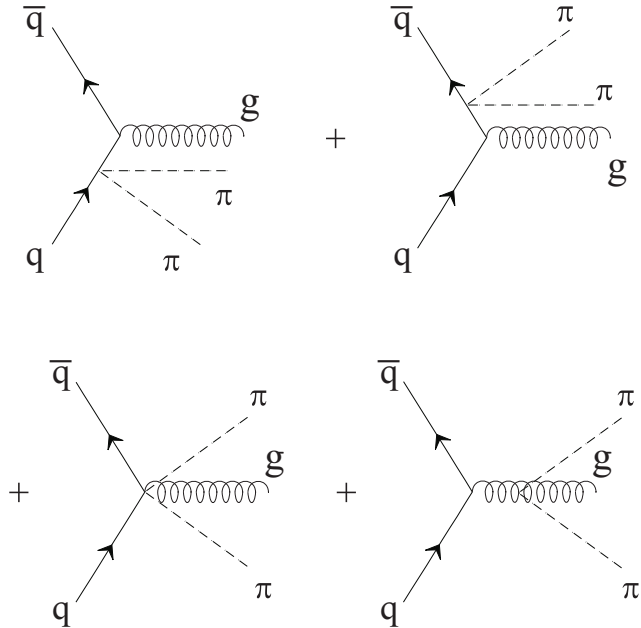}
\caption{\small Typical Feynman diagrams for production of monojet plus
branons.  This is taken from Ref.~\protect\cite{branon-3}.\label{branon-fig1}}
\end{figure}

\begin{figure}[th!]
\centering
\includegraphics[width=4in]{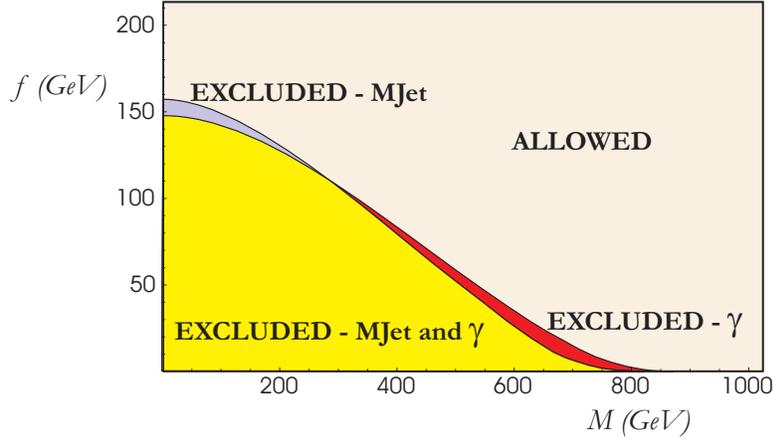}
\caption{\small Exclusion limit of branon parameter space due to
monojet and monophoton plus missing energies..
 This is taken from Ref.~\protect\cite{branon-3}.\label{branon-fig2}}
\end{figure}

{\it Branons.} 
So far in the brane world scenario the branes are assumed rigid, i.e., there 
are violations of momentum conservation along the transverse directions.
However, in reality the brane could be flexible, i.e., it has tension.
In a lot of calculations that involve intermediate graviton KK state 
exchanges, e.g., $e^+ e^- \to G^{(n)} \to \gamma\gamma, \ell^+ \ell^-$,
we encounter this kind of sum
\begin{equation}
\label{branon-sum}
g^2 \sum_k \frac{1}{p^2 - k^2/R^2}\;,
\end{equation}
where $m_k =k/R$ is the mass of the graviton KK states.  This sum in fact
diverges when the number of extra dimensions is 2 or more.  Thus, in the
actual calculation one has to impose an upper cutoff in the sum.  Typically,
the cutoff is at the scale $M_S$.  It becomes an effective theory below
the scale $M_S$.
In the consideration of a flexible brane, an exchange of a KK graviton
means a ``deformation'' of the brane due to momentum conservation.  The
higher the KK state exchange, the larger the deformation becomes.  
Large deformation, just like stretching a string, requires more energy.
Therefore, exchanges of higher KK states would suffer stronger suppression.  
Bando {\it et al.} \cite{branon-1} modified the coupling 
\[
g \longrightarrow g_k = g \exp\left( - \frac{k^2}{R^2} \frac{M_S^4}{f^4} 
\right ) \;,
\]
where $f$ is the tension of the brane, to represent the suppression on 
the exchange of the KK state with mass $k/R$.
 Thus, the sum in 
Eq.~(\ref{branon-sum}) can extend to infinity without a cutoff \cite{branon-1}.
In view of such an interesting observation, Cembranos {\it et al.} 
\cite{branon-2,branon-3} interpreted the deformation as branon.  The 
branon interacts with the energy-momentum tensor of the SM particles
\begin{equation}
{\cal L}_{\rm int} = \frac{1}{8 f^4} \left( 4 \partial_\mu \pi^\alpha
 \partial_\nu \pi^\alpha - M^2 \pi^\alpha \pi^\alpha g_{\mu\nu} \right )
\; T^{\mu\nu}_{\rm SM} \;.
\end{equation}
Note that the parity of the brane requires branons to couple pairwise 
to SM particles, so that the lightest branon is stable.  The lightest 
branon is then a possible dark matter candidate \cite{branon-2}.
The collider signature for branons would be missing energies.
Cembranos {\it et al.} did a study of branons at hadron colliders, in 
particular the monojet plus missing energy signal.  Typical Feynman 
diagrams for the process are shown in Fig. \ref{branon-fig1}.
Using existing data on monojet and monophoton plus missing energies certain 
parameter space of the branon interaction can be ruled out, as shown in
Fig.~\ref{branon-fig2} \cite{branon-3}.  
L3 Coll. \cite{branon-4} also searched for branons using 
$e^+ e^- \to \pi \pi Z, \pi\pi q\bar q, \pi \pi \gamma$.  Branon mass
below 103 GeV with small brane tension is excluded.

\subsection{Trans-Planckian}

\begin{figure}[th!]
\centering
\includegraphics[width=5.5in]{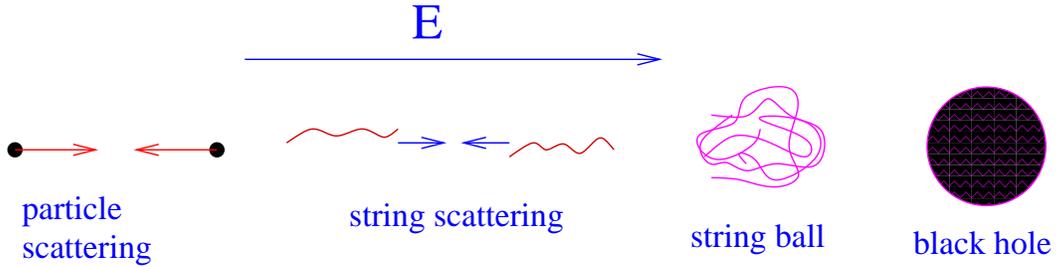}
\caption{Figure showing the trans-Planckian signatures.
\label{bh}}
\end{figure}

Since the fundamental Planck scale is at TeV, so at the future
hadronic collider, the LHC, the energy can surpass the fundamental
Planck scale.  Particle scattering would show the features of
quantum gravity \cite{fischler,hole,scott,greg}, 
because the fundamental Planck scale is at which
the quantum gravity effects become strong.  The cartoon in Fig.
\ref{bh} shows the behavior of the scattering when the energy
scale is getting higher and higher.  When it approaches the string
scale, the scattering is characterized by string scattering.  As
energy further increases, the string becomes highly excited and
entangled like a string ball \cite{sb}.  Eventually, the energy reaches a
transition point and everything will turn into a black hole.

\begin{figure}[th!]
\centering
\includegraphics[width=4.5in]{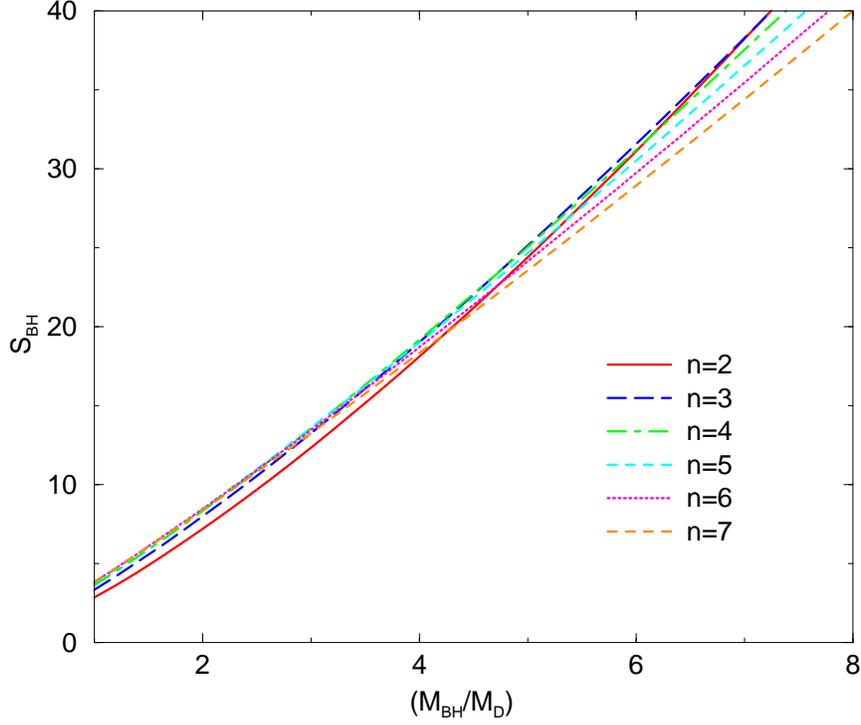}
\caption{\small Entropy of a BH vs its mass. From
Ref.~\protect\cite{km-2}. \label{s-bh} }
\end{figure}

A black hole (BH) is characterized by its mass, charge, and
angular momentum.  Here we simply look at the uncharged and
nonrotating BH.  The Schwarzchild radius and entropy of a BH with
a mass $M_{\rm BH}$ in $n+4$ dimensions are given by,
respectively, \cite{myer}
\begin{eqnarray}
 R_{\rm BH} &=& \frac{1}{M_D}\; \left ( \frac{M_{\rm BH}}{M_D}
\right)^{\frac{1}{n+1}}\; \left( \frac{ 2^n \pi^{ \frac{n-3}{2}}
\Gamma(\frac{n+3}{2} )}{n+2} \right )^{\frac{1}{n+1}} \\
 S_{\rm BH} &=& \frac{4\pi}{n+2}\; \left ( \frac{M_{\rm BH}}{M_D}
\right)^{\frac{n+2}{n+1}}\; \left( \frac{ 2^n \pi^{ \frac{n-3}{2}}
\Gamma(\frac{n+3}{2} )}{n+2} \right )^{\frac{1}{n+1}} \;.
\end{eqnarray}
As argued in a number of papers \cite{giddings,km}, 
the entropy of the BH must be
large in order that the object is in fact a BH.  Here we follow
the convention that the entropy $S_{\rm BH} \agt 25$.  We show the
entropy $S_{\rm BH}$ vs the mass of the BH in Fig. \ref{s-bh}. It
is clear that the entropy increases with the mass, and the
requirement of $S_{\rm BH}\agt 25$ is roughly equivalent to
$M_{\rm BH}> 5 M_D$.

The production cross section of a BH in a collision is given
by \cite{scott,greg}
\begin{equation}
\sigma = \pi R_{\rm BH}^2 \;,
\end{equation}
which is based on a naive semi-classical argument.  Suppose the
two incoming particles involved in the collision have a
center-of-mass energy $\sqrt{s}$ and they want to coalesce into a
BH, they can only do so if they are within the event horizon of
the BH to be produced.  Therefore, they can only produce a BH if
their impact parameter is less than the radius, and
thus resulting in the cross section formula above. The decay of
the BH is somewhat complicated.  Naively, one would expect the BH,
as a quantum gravity object, would decay into gravitons, which
then go to the extra dimensions and get lost.  Experimentally, it
would not be seen.  However, the work by Emparan, Horowitz, and
Myers \cite{emp} showed that it is not the case.  Since the main phase of the
BH decay is via the Hawking radiation, the wavelength
corresponding to the Hawking temperature is much larger than the
$R_{\rm BH}$.  Thus, the BH behaves like a $s$-wave point source
and decays equally into the brane and bulk modes.  Since in the
setup there is only one graviton in the extra dimensions, but all
SM particles on the brane, so the BH decays most of the time into
the brane particles,
\footnote{
For alternative viewpoints on BH decays, see Refs. \cite{bh-decay}.
}
 i.e., the SM particles, and it could be
observed in experiments.  The BH decays ``blindly" according to the
particle degrees of freedom.  The ratio
\[
Z, W, H, \gamma, g;\;\; u, d, s, c, b, t;\;\; e, \mu, \tau, \nu_e,
\nu_\mu, \nu_\tau =30:72:18
\]
and  the ratio of hadronic:leptonic $\sim 5:1$. In addition, a
nonrotating BH decays isothermally, and so for a BH of a few TeV
it decays into $30-50$ particles, each of which then has an energy
of a few hundred GeV.  Therefore, the BH event will look like a
spherical fireball \cite{km,km-2}.  
Such events are very clean and suffer from no
background in collider experiments.  One can do the event
counting.  We give the estimates for the BH production cross
section at the LHC in Table \ref{table2} \cite{km,km-2}.

\begin{table}[th!]
\centering
\caption{ \label{table2} Cross section in pb for BH
production at the LHC.  Here we included the contributions for the
$2\to 1$ and $2\to 2$ subprocesses, and $y \equiv M_{\rm BH}/M_D$. }
\begin{tabular}{|c|lll|}
\hline
  &  $n=4$ & $n=5$ &  $n=6$ \\
\hline
\underline{$M_D=1.5$ TeV} &       &       &  \\
  $y=1$                   & 9200   & 13000  & 18000 \\
  $y=2$                   & 890    & 1250   & 1600   \\
  $y=3$                   & 110    &  150   & 190  \\
  $y=4$                   & 12     &  15    & 21 \\
  $y=5$                   & 0.99   &   1.3  & 1.6  \\
\hline
\underline{$M_D=3$ TeV}  &  &   &  \\
  $y=1$                   & 179   & 240  & 330\\
  $y=2$                   & 2.3 & 3.2  &  4.3\\
  $y=3$                   & 0.0085& 0.011 & 0.015\\
  $y=4$                   & $2.6\times 10^{-7}$ & $3.5\times 10^{-7}$ &
                  $4.5\times 10^{-7}$  \\
  $y=5$                   &  -& -  &  - \\
\hline
\end{tabular}
\end{table}

Other trans-Planckian objects include string balls \cite{sb} 
and $p$-branes \cite{p-brane}.
Dimopoulos and Emparan \cite{sb} pointed out that when a BH reaches a
minimum mass, it will transit into a state of highly excited and
jagged strings -- string balls (SB), the transition point is at
\[
M_{\rm BH}^{\rm min} = M_s/g_s^2 \;,
\]
where $M_s$ is the string scale and $g_s$ is the string coupling.
SBs can be considered the stringy progenitors of BHs. The
correspondence principle states that the properties of a BH with a
mass $M_{\rm BH}$ match those of a string ball of a string theory
with $M_s/g_s^2= M_{\rm BH}$.   Thus, their production cross
section at the transition point should match, i.e.,
\begin{equation}
\left. \sigma(SB) \right|_{M_{SB} = M_s/g_s^2} = \left. \sigma(BH)
\right|_{M_{BH} = M_s/g_s^2} \;.
\end{equation}

We can parameterize the production cross section of the SB as
follows. When the energy is above $M_s$ but below $M_s/g_s$, the
scattering is described by string-string scattering, the amplitude
of which should scale as $\sim \hat s/M_s^4$.  When the energy
reaches  $M_s/g_s$, saturation of unitarity sets $\sigma$ to be a
constant until it hits the correspondence point, after which the
SB production cross section is replaced by the BH cross section.
Thus, we have the following for the SB/BH production:
\begin{equation}
\hat\sigma ({\rm SB/BH}) = \left \{ \begin{array}{ll}
\frac{\pi}{M_D^2}\; \left ( \frac{M_{\rm BH}}{M_D}
\right)^{\frac{2}{n+1}}  \left[ f(n) \right ]^2 &
                         \frac{M_s}{g_s^2} \le M_{\rm BH} \\
& \\
\frac{\pi}{M_D^2}\; \left ( \frac{M_s/g_s^2}{M_D}
\right)^{\frac{2}{n+1}}  \left[ f(n) \right ]^2
 = \frac{ \pi}{M_s^2}  \left[ f(n) \right ]^2 &
                \frac{M_s}{g_s} \le M_{\rm SB} \le \frac{M_s}{g_s^2} \\
&\\
\frac{ \pi g_s^2 M_{\rm SB}^2 }{M_s^4}  \left[ f(n) \right ]^2 &
                M_s \ll M_{\rm SB} \le \frac{M_s}{g_s}
\end{array}
 \right.
\end{equation}
where 
\[
f(n) = \left( \frac{ 2^n \pi^{ \frac{n-3}{2}}
\Gamma(\frac{n+3}{2} )}{n+2} \right )^{\frac{1}{n+1}} 
\]
The decay of a SB would be similar to the BH, and thus most of the
time into the SM particles.

A $p$-brane is a solution to the Einstein equation in multi
dimensions.  A BH is considered a $0$-brane, therefore $p$-branes,
in principle, can also be produced in hadronic collisions.
Consider an uncharged, static $p$-brane with mass $M_{p \rm B}$.
The $p$-brane wraps on $r (\le m)$ small extra dimensions and on
$p-r (\le n-m)$ large extra dim. 
\footnote{
The configuration has $n$ total extra dimensions, of which 
$m$ dimensions are small ($\sim 1/M_*$) and $n-m$ are large ($\gg 1/M_*$).
Here $M_*$ is another notation in literature for the fundamental Planck
scale, $M_* = M_S$.}
The radius of the $p$-brane is
given by
\begin{equation}
R_{p \rm B} = \frac{1}{\sqrt{\pi} M_*} \, \gamma(n,p) \, V_{p\rm
B}^{ \frac{1}{1+n-p} } \, \left( \frac{M_{p\rm B}}{M_*} \right)^{
\frac{1}{1+n-p} }
\end{equation}
where
\begin{eqnarray}
 V_{p \rm B} &=& l_{n-m}^{p-r} \, l_m^{r}
\approx \left( \frac{M_{\rm Pl}}{M_*} \right )^{
\frac{2(p-r)}{n-m} } \;, \nonumber \\
\gamma(n,p) &=& \left[ 8 \Gamma \left( \frac{3+n-p}{2} \right)
\sqrt{ \frac{1+p}{(n+2)(2+n-p)} } \right ]^{\frac{1}{1+n-p} }
\nonumber \;.
\end{eqnarray}
Note that $R_{p\rm B} \to R_{\rm BH}$ in the limit $p=0$. The
production cross section of the $p$-brane is similar to the BH,
given by
\begin{equation}
\hat \sigma (M_{p\rm B}) = \pi R^2_{p\rm B} \;.
\end{equation}
The radius $R_{p\rm B}$ of a $p$-brane is suppressed by some
powers of the volume $V_{p\rm B}$ wrapped by the $p$-brane. In
order to achieve the maximum cross section, the value of $V_{p\rm
B}$ should be minimum, which occurs when the $p$-brane wraps
entirely on the small extra dimensions only, i.e., $r=p$. The
ratio of $p$-brane cross section to BH cross section is given by
\begin{eqnarray}
R &\equiv& \frac{\hat \sigma (M_{p\rm B}=M)}{\hat \sigma (M_{\rm
BH}=M)} \nonumber \\
&=& \left ( \frac{M_*}{M_{\rm Pl}}
\right)^{\frac{4(p-r)}{(n-m)(1+n-p)}}  \left(\frac{M}{M_*}
\right )^{ \frac{2p}{(1+n)(1+n-p)} }  \left(
\frac{\gamma(n,p)}{\gamma(n,0)} \right )^2
\end{eqnarray}

\begin{figure}[th!]
\centering
 \includegraphics[width=4.5in]{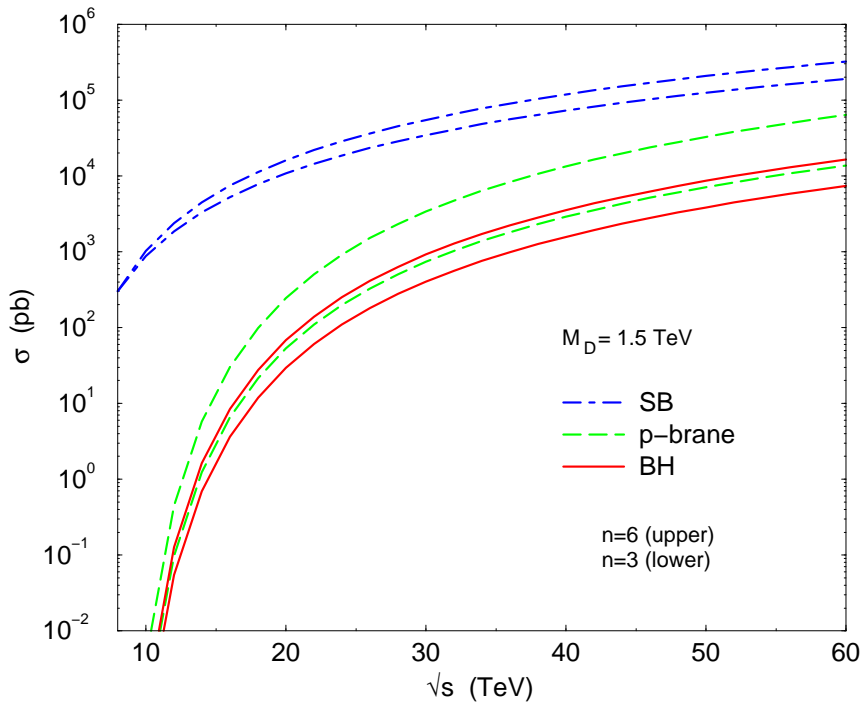}
\caption {\small \label{cross} Production cross sections for black
holes, string balls, and $p$-branes.  We have chosen $M_D=1.5$
TeV, $M_{\rm BH}^{\rm min}=5 M_D$, and $M_{\rm SB}^{\rm min}=2
M_s$.  From Ref.~\protect\cite{km-2}. }
\end{figure}

A comparison of production cross sections of BHs, SBs, and $p$-branes
is shown in Fig. \ref{cross}.  Since the production threshold of SBs is
much lower than BHs, SB cross section is therefore much larger.  The
$p$-brane cross section is somewhat larger than the BHs.

\begin{figure}[th!]
\centering
\includegraphics[width=3.3in]{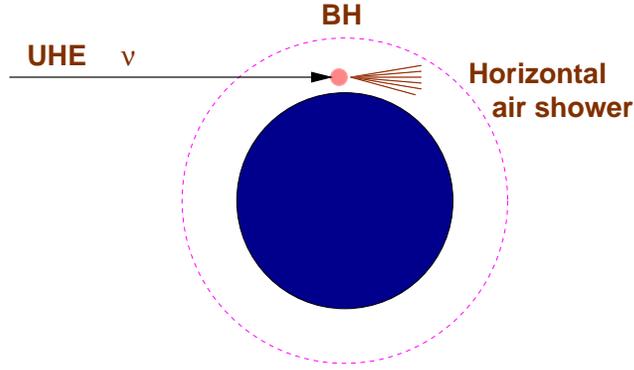}
\caption{\label{bh-cr} Cartoon showing that the UHECR neutrino
produces a BH and gives a horizontal air shower.}
\end{figure}

\begin{figure}[th!]
\centering
\includegraphics[width=4in]{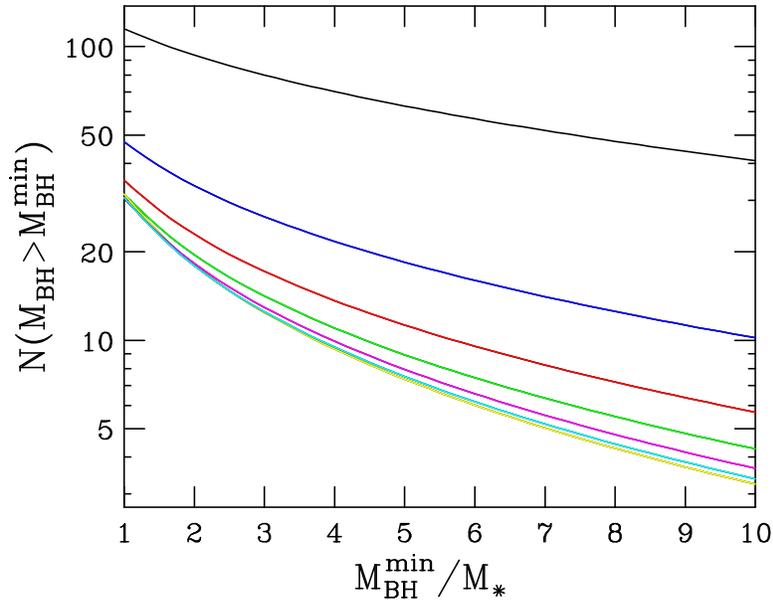}
\caption{\label{auger} Number of BH events detected by the
ground array in 5 Auger site-years for $n=1-7$ from above.  This is taken
from Ref.~\protect\cite{feng}. }
\end{figure}

Feng and Shapere \cite{feng} pointed out another possibility of observing the
BH in the ultra-high energy cosmic ray (UHECR) experiments.  The
UHECR is the beam while our atmosphere is the target.  The UHECR
has a neutrino component that can penetrate deeply into the
atmosphere without interacting. It is the neutrino component in the
UHECR that produces the distinct signature for BHs.  The largest
chance that a neutrino can interact with the nucleons in the
atmosphere to produce a BH is when it traverses horizontally
across the atmosphere (shown in the cartoon of Fig. \ref{bh-cr}.) The
BH then decays instantaneously, thus producing a horizontal air
shower.  Such deeply penetrating horizontal air showers are to be
counted and compare with the SM prediction.  The number of BH
events expected for the run at the Pierre-Auger Observatory is
shown in Fig. \ref{auger}.  A partial list of works in this area
are listed in Refs. \cite{bh-all}.

\section{Randall-Sundrum model}

\begin{figure}[th!]
\centering
\includegraphics[width=4in]{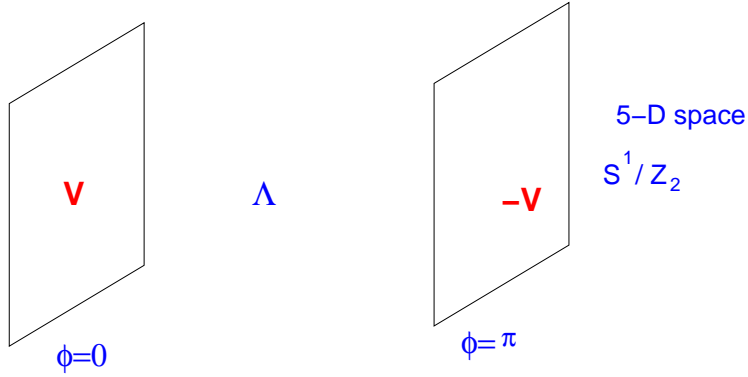}
\caption{\label{rs} The setup in the Randall-Sundrum
model.}
\end{figure}

The Randall-Sundrum (RS) model \cite{RS} beautifully explains the
gauge hierarchy with a moderate number through the exponential.
The setup of the branes and the bulk is shown in Fig. \ref{rs}.

Just like the ADD model, since only graviton propagates in the
extra dimensions, only gravity can probe the extra dimensions.
However, the graviton KK states are very different from the ADD
model.  The most distinct feature of the RS model is the unevenly
spaced KK spectrum for the gravitons, namely, proportional to the
zeros of the $n$th modified Bessel function \cite{RS-1}.  Phenomenology
associated with the modulus field (known as the radion),
describing the fluctuation in the separation of the two branes, is
another interesting feature of the RS model \cite{RS-2,RS-3}.  
There is also the
possibility that the radion can mix with the Higgs boson \cite{RS-4}.

\subsection{Graviton}
The graviton field can be obtained by fluctuation of the metric
\begin{equation}
 G_{\alpha\beta} = e^{-2ky} \eta_{\alpha\beta} + 2
h_{\alpha\beta}/M_5^{3/2} \;.
\end{equation}
After compactification, the KK states of the graviton has the
spectrum given by
\begin{equation}
m_n = x_n \, \frac{\Lambda_\pi}{\sqrt{2}}\, 
\frac{k}{\overline M_{\rm Pl}}
\end{equation}
where $x_n$ is the zero of the $n$-th modified Bessel function.
Numerically, $x_1, x_2, x_3 =3.83, 7.02, 10.17$, respectively.  
Note that the spectrum is very
different from that of flat metric. The interactions are given by
\begin{equation}
{\cal L} = -\frac{1}{\overline{ M_{\rm Pl}}} T^{\mu\nu}
h^{(0)}_{\mu\nu} - \frac{1}{\Lambda_\pi} T^{\mu\nu}(x)
\sum_{n=1}^\infty h^{(n)}_{\mu\nu}(x) \;,
\end{equation}
from which  we can see that the zeroth mode essentially decouples
because the coupling is suppressed by $1/M_{\rm Pl}$ while the
KK states have a coupling strength of $1/\Lambda_\pi$. The
phenomenology of the RS model is very different from the ADD model
in two aspects: (i) the spectrum of the graviton KK states are
discrete and unevenly spaced while it is uniform, evenly spaced,
and effectively a continuous spectrum in the ADD model,  and (ii)
each resonance in the RS model has a coupling strength of $1/$TeV
while in the ADD model only the collective strength of all
graviton KK states gives a coupling strength of $1/$TeV.

\begin{figure}[th!]
\centering
\includegraphics[angle=90,width=4.5in]{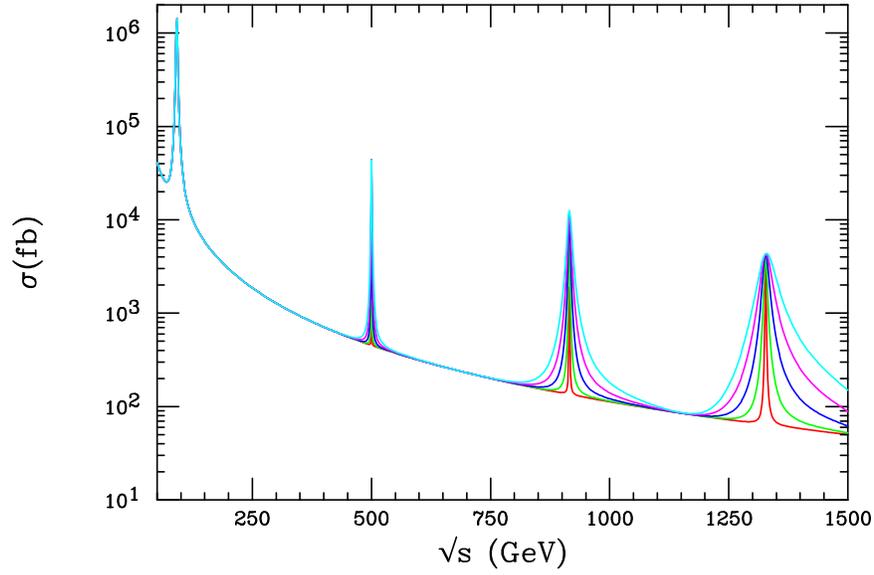}
\caption{\label{rs-1} Cross section of $e^+ e^- \to \mu^+
\mu^-$ vs $\sqrt{s}$, including the effects of the RS graviton KK
states.  This is taken from Ref.~\protect\cite{review1}.}
\end{figure}

\begin{figure}[th!]
\centering
\includegraphics[angle=90,width=3.3in]{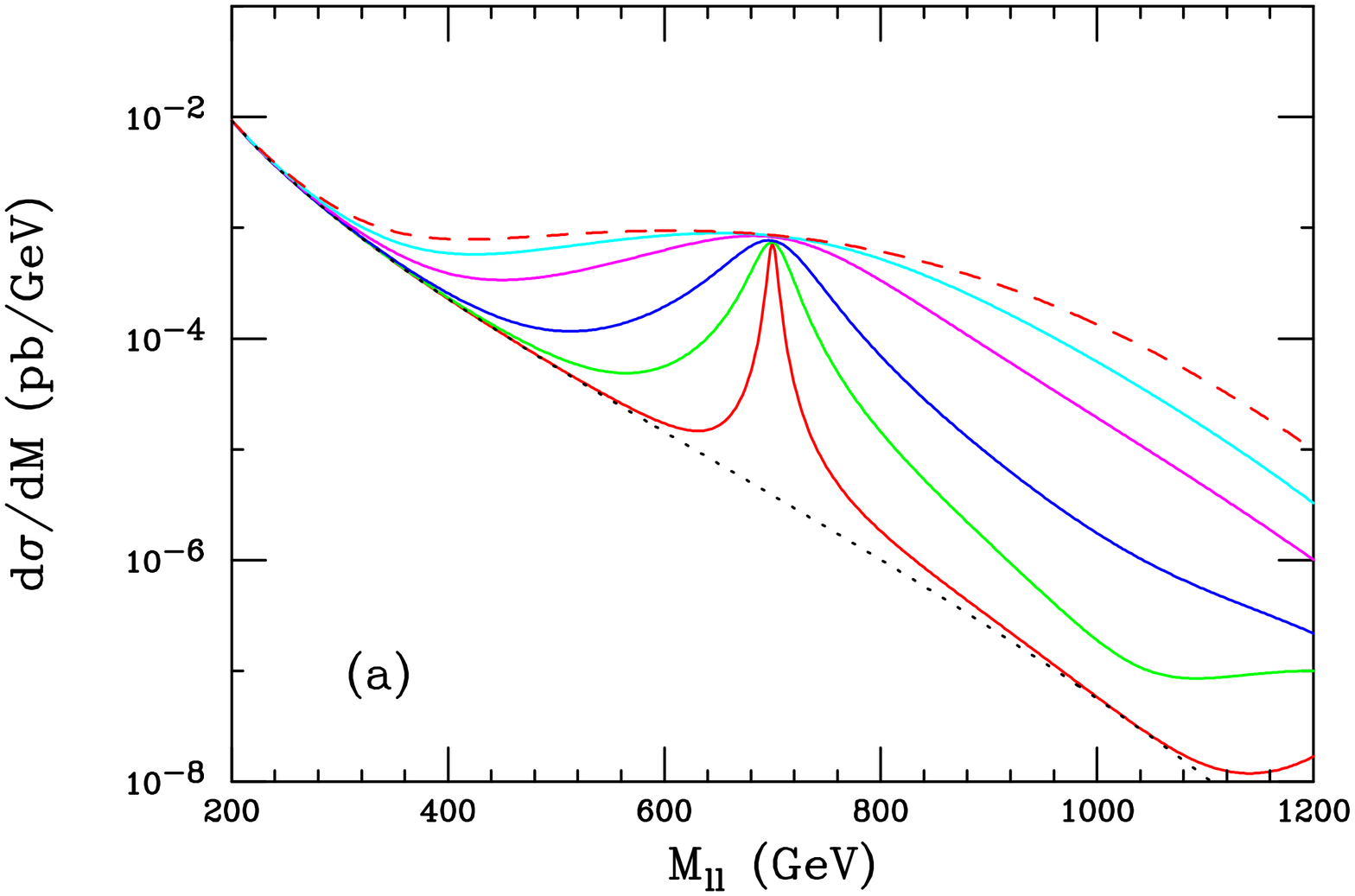}
\includegraphics[angle=90,width=3.3in]{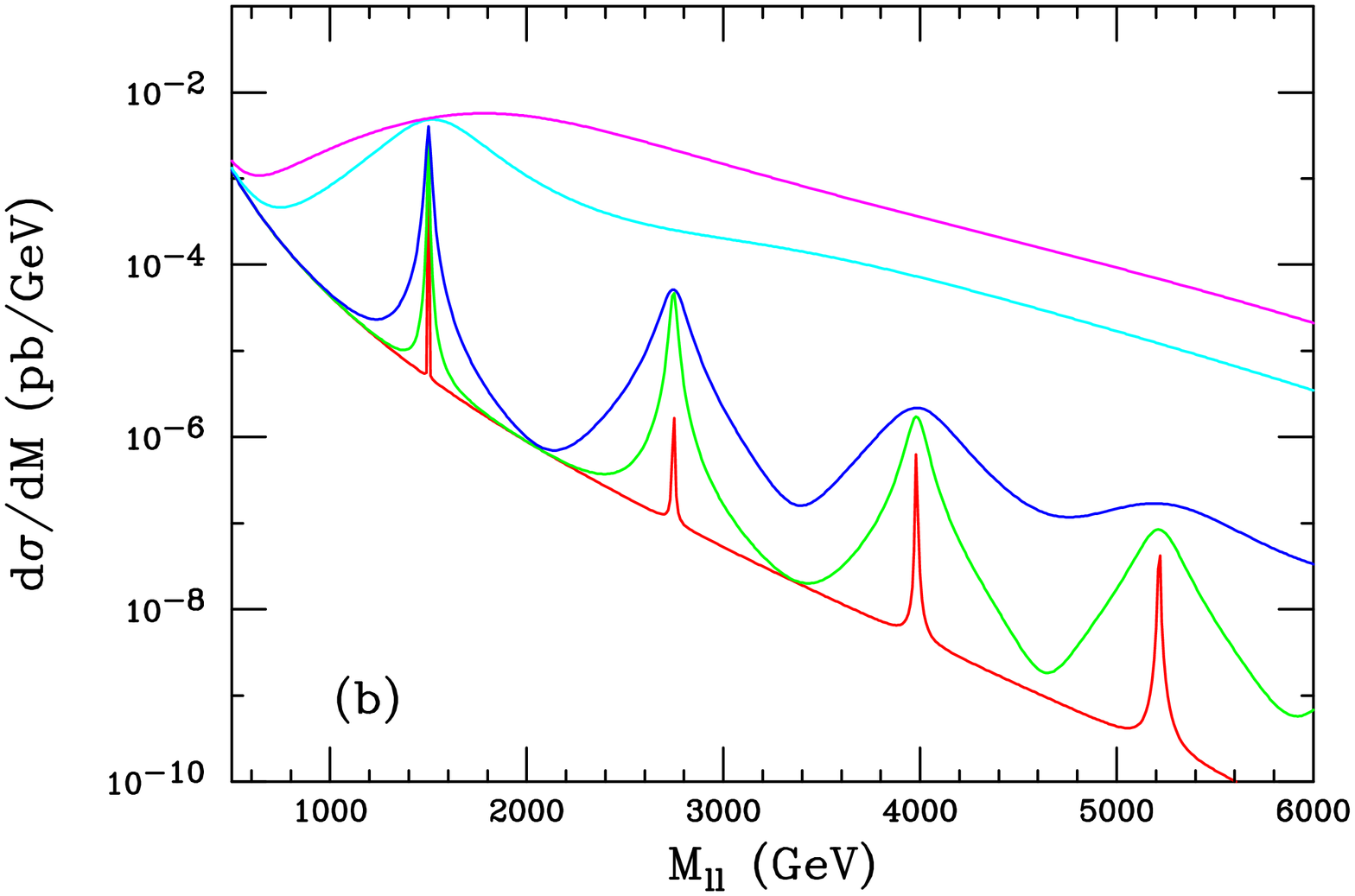}
\caption{\label{rs-2} The invariant mass distribution for
the Drell-Yan process at the (a) Tevatron and (b) the LHC. This is taken
from Ref.~\protect\cite{RS-5}.}
\end{figure}

Figure \ref{rs-1} shows the resonance spectrum in the
channel $e^+ e^- \to \mu^+ \mu^-$.  The resonance spectrum clearly
indicates that it is a discrete one and is unevenly spaced. The
best present limit comes from the Drelly-Yan production at the
Tevatron.  The effects of the graviton KK states on the Drell-Yan
production at the Tevatron and at the LHC are summarized in Fig.
\ref{rs-2}.  Davoudiasl et al. \cite{RS-1,RS-5} 
showed that the present Drell-Yan
data can rule out a portion of the parameter space of the RS
model.  

There are some updates on the search for RS graviton at the Tevatron 
\cite{kajf}.
CDF has been using the Drell-Yan data from electron and muon samples and
diphoton data to constrain the RS graviton.  The limit is $M_{G^{(1)}}
> 605-690$ GeV for $k/M_{\rm Pl}=0.1$ \cite{kajf}. 
On the other hand, D\O\ used dielectromagnetic objects 
(dielectron and diphoton) and obtained a limit $M_{G^{(1)}}
> 785$ GeV for $k/M_{\rm Pl}=0.1$ \cite{kajf}.

\subsection{Radion}
The RS model has a 4D massless scalar, radion, describing the
fluctuation in the background metric
\[
ds^2 = e^{-2 k \phi T(x)} g_{\mu\nu}(x) \, dx^\mu dx^\nu - T^2(x)
d\phi^2
\]
where $T(x)$ is the modulus field (radion) describing the distance
between the two branes.  Since the gauge hierarchy is explained
by a particular brane separation, a stabilization mechanism is
necessary to achieve that.  Goldberger and Wise \cite{RS-2,RS-3} 
used a bulk scalar
field to generate a potential, and the modulus field acquires a
$O(0.1-1 TeV)$ mass with a coupling strength $1/$TeV.

The interactions of the radion with the SM particles are given by
\begin{equation}
{\cal L}_{\rm int} = \frac{\phi}{\Lambda_\phi} \; T^\mu_\mu ({\rm
SM}) \;,
\end{equation}
where $\Lambda_\phi= \langle \phi \rangle$ is of order TeV and
\begin{eqnarray}
T^\mu_\mu ({\rm SM}) &=& \sum_f m_f \bar f f - 2 m_W^2 W_\mu^+
W^{-\mu} -m_Z^2 Z_\mu Z^\mu \nonumber \\
&& + (2m_h^2 h^2 - \partial_\mu h
\partial^\mu h  ) + ... \;.
\end{eqnarray}
It is clear that the interactions are very similar to those of the
SM Higgs boson with the replacement of the vacuum expectation value.
However, the radion has anomalous couplings from the trace anomaly
to a pair of gluons (photons), in addition to the loop diagrams
with the top-quark (the top-quark and $W$ boson):
\[
T^\mu_\mu({\rm SM})^{\rm anom} = \sum_a \frac{\beta_a (g_a)}{2g_a}
F_{\mu\nu}^a F^{a \mu\nu} \;,
\]
where
\[
\beta_{\rm QCD}/2g_s = -(\alpha_s/8\pi) b_{\rm QCD}\;\;\; {\rm
and}\;\;\; b_{\rm QCD} = 11 - 2 n_f/3
\]
and
\[
\beta_{\rm QED}/2e = -({\alpha/8\pi}) b_{\rm QED}\;\;\; {\rm and}
\;\;\; b_{\rm QED} = -11/3.
\]

\begin{figure}[th!]
\centering
\includegraphics[width=4.5in]{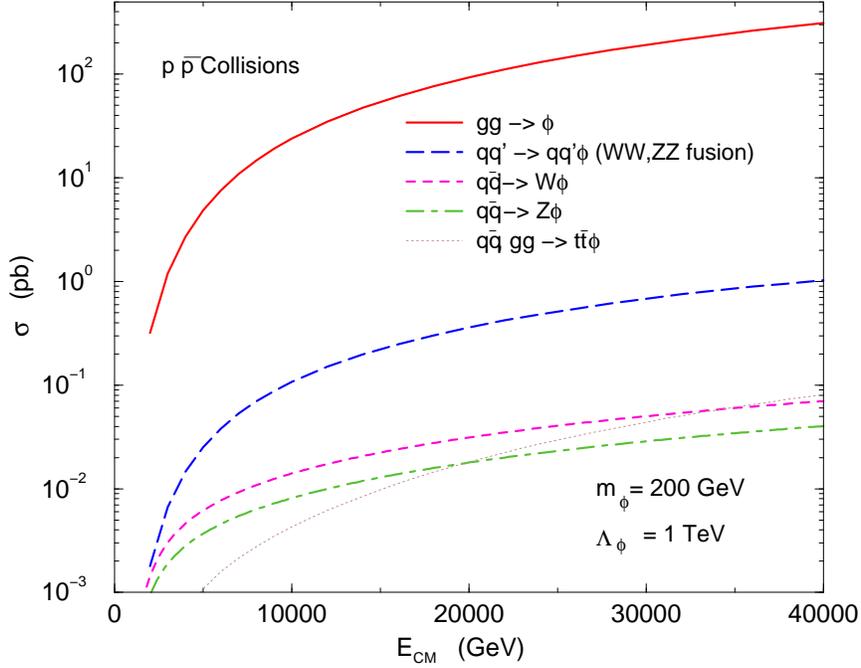}
\caption{ \label{rs-5} Hadronic production cross sections
for the radion. From Ref.~\protect\cite{RS-6}.}
\end{figure}

Because of the anomalous coupling of the radion to gluons, the
gluon fusion will be the most important production channel for the
radion in hadronic collisions, followed by the $WW,ZZ$ fusion.
Figure \ref{rs-5} shows the production cross section of the radion
at $p\bar p$ colliders.

\begin{figure}[th!]
\centering
\includegraphics[width=4.5in]{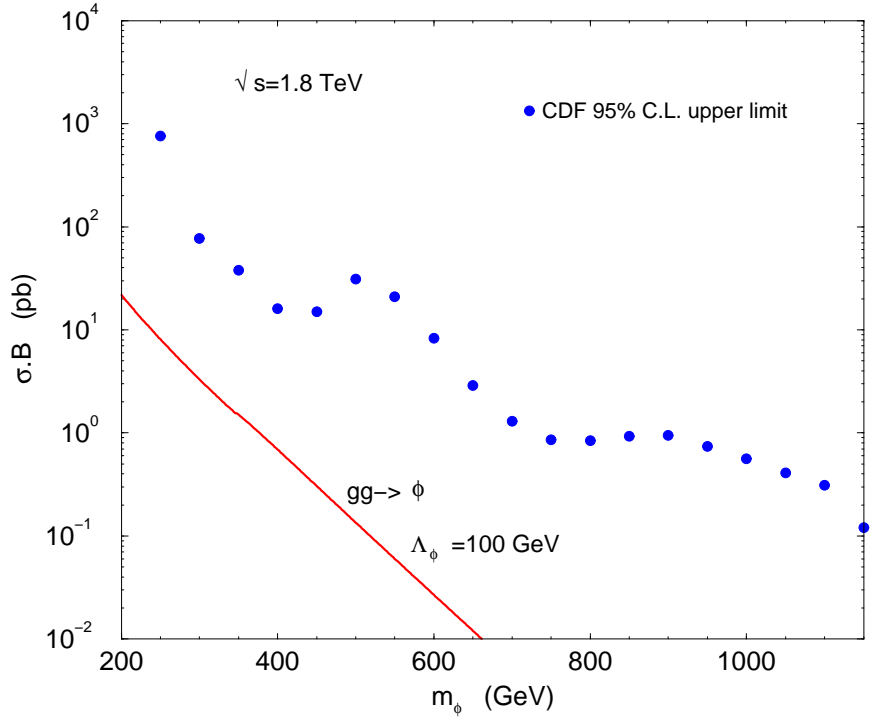}
 \caption{ \label{rs-6}95\% C.L. upper limit on $\sigma
 \cdot B$ for a hypothetical resonance decaying into dijet. From 
Ref.~\protect\cite{RS-6}. }
 \end{figure}

\begin{figure}[th!]
 \centering
 \includegraphics[width=4.5in]{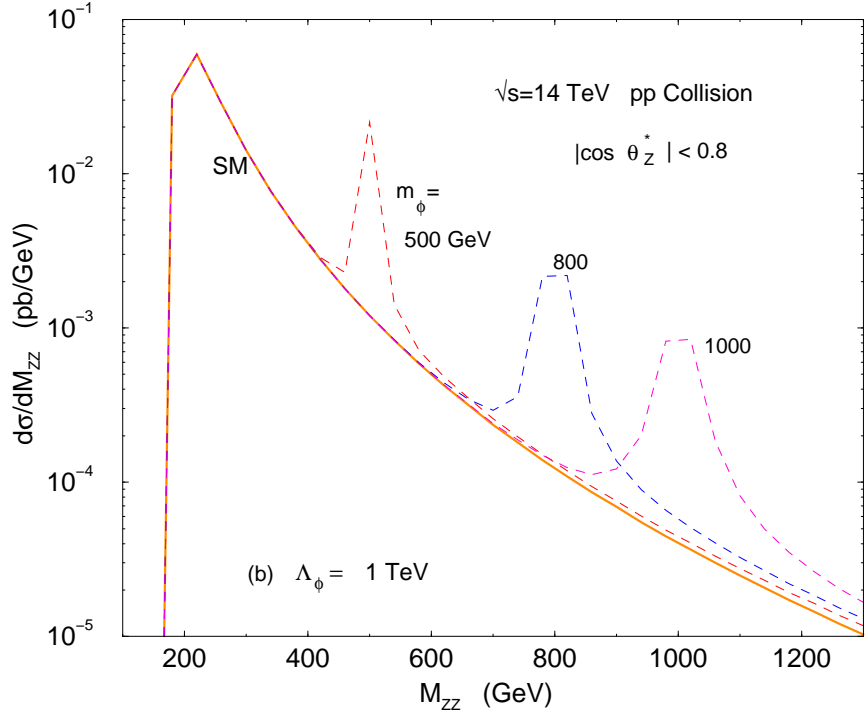}
\caption{ \label{rs-7}The invariant mass $M_{ZZ}$ spectrum
for the radion production.  From Ref.~\protect\cite{RS-6}. }
\end{figure}

One may expect that since the gluon fusion is extraordinary large,
the Tevatron dijet data might have some restriction on the radion.
Figure \ref{rs-6} shows the 95\% C.L. upper limit on the $\sigma
\cdot B$, which is the cross section from a hypothetical massive
particle times the branching ratio into dijet.  However, the
production cross section of the radion is below the CDF curve, and
thus no limit is placed on the radion.  Therefore, it is not desirable 
to detect the radion through its dijet decay mode because of
the large QCD background.  On the other hand, for a heavy radion
the $ZZ$ decay mode opens up and the detection is a golden mode,
similar to the SM Higgs boson.  Figure \ref{rs-7} shows the
invariant mass spectrum for $ZZ$ production.  It is clear that the
radion peak is very distinct above the background.

\subsection{Radion-Higgs mixing}
Since both the gauge and Poincare invariance do not forbid the
mixing between the gravity scalar and the Higgs boson, one should
expect that, in general, they should mix.  The mixing term in
the action is given by \cite{RS-4,RS-7}
\begin{equation}
S_\xi=\xi \int d^4 x \sqrt{g_{\rm vis}}R(g_{\rm vis})\hat
H^\dagger \hat H\,,
\end{equation}
where $R(g_{\rm vis})$ is the Ricci scalar for the induced metric
on the visible brane, and $\xi \to 0$ in the limit of no mixing.
The free Lagrangian of the Higgs and radion is \cite{RS-7}
\begin{eqnarray}
{\cal L}_0&=& -\frac{1}{2}\left\{1+6\gamma^2 \xi
\right\}\phi_0\Box\phi_0   -\frac{1}{2} \phi_0
m_{\phi_0}^2\phi_0  \nonumber \\
&& -\frac{1}{2}
 h_0 (\Box+m_{h_0}^2)h_0-6\gamma \xi \phi_0\Box h_0 \;.
\end{eqnarray}
Physical states $h$ and $\phi$ can be introduced to diagonalize
${\cal L}_0$, defined by

\begin{eqnarray}
\left(
\begin{array}{c}
  h_0 \\
  \phi_0 \\
\end{array}
\right) &=& \left(
\begin{array}{cc}
  1 & 6 \xi \gamma/Z \\
  0 & -1/Z \\
\end{array}
\right) \left(
\begin{array}{rc}
  \cos\theta & \sin\theta \\
  -\sin\theta & \cos\theta \\
\end{array}
\right) \left(\begin{array}{c}
  h \\
  \phi \\
\end{array}
\right) \\ \label{def-matrix} &\equiv& \left(
\begin{array}{cc}
  d & c \\
  b & a \\
\end{array}
\right) \left(\begin{array}{c}
  h \\
  \phi \\
\end{array}
\right) \,,
\end{eqnarray}
where
\[
Z^2\equiv 1+6\xi\gamma^2(1-6\xi)\equiv \beta-36\xi^2\gamma^2\,.
\]

A nonzero $\xi$ will induce some triple couplings \cite{RS-7,RS-8}
\[
h\,\mbox{-}\,\phi\,\mbox{-}\,\phi, \quad
h_{\mu\nu}^{(n)}\,\mbox{-}\,h\,\mbox{-}\,\phi,\quad
\phi\,\mbox{-}\,\phi\,\mbox{-}\,\phi,\quad
h_{\mu\nu}^{(n)}\,\mbox{-}\,\phi\,\mbox{-}\,\phi \;.
\]
All phenomenological signatures of the RS model including the
radion-Higgs mixing are specified by five parameters
\begin{equation}
\label{parameter} \xi,\quad \Lambda_\phi,\quad \frac{m_0}{M_{\rm
Pl}},\quad m_\phi,\quad m_h \,,
\end{equation}
which in turns determine $\Lambda_W$ and KK graviton masses
$m_G^{(n)}$ as
\begin{equation} \label{parameter-relation}
\Lambda_W = \frac{\Lambda_\phi}{\sqrt{3}} , \quad m_G^{(n)} = x_n
\frac{m_0}{M_{\rm Pl}} \frac{\Lambda_W}{\sqrt{2}} \,.
\end{equation}

\begin{figure}[th!]
\centering
\includegraphics[angle=90,width=3.3in]{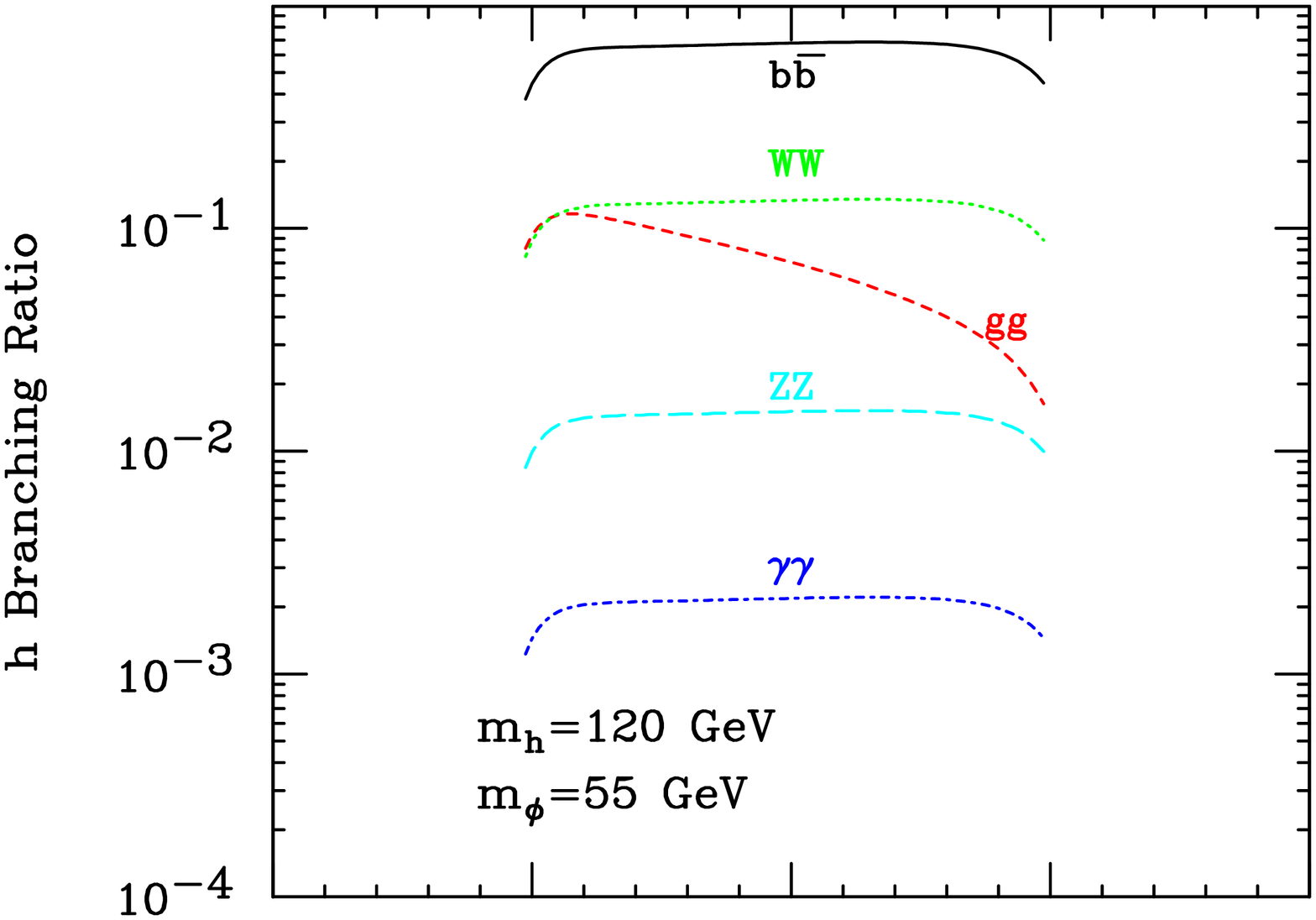}
\includegraphics[angle=90,width=3.3in]{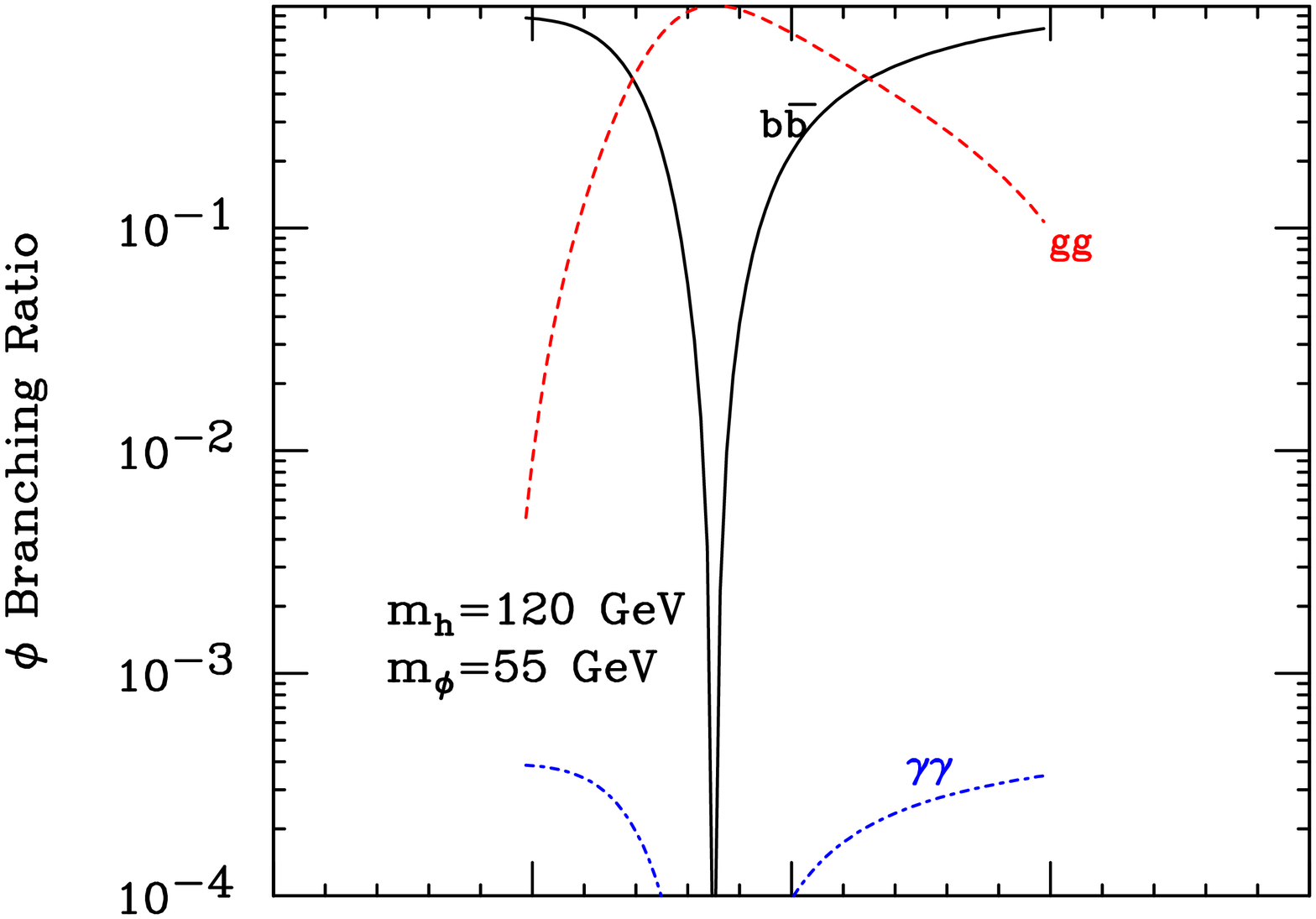}
\caption{\label{rs-8}
The effect of the radion-Higgs mixing
on the branching ratios of (a) the Higgs and (b) the radion. This is 
taken from Ref.~\protect\cite{RS-7}.}
\end{figure}

\begin{figure}[th!]
\centering
\includegraphics[width=3in]{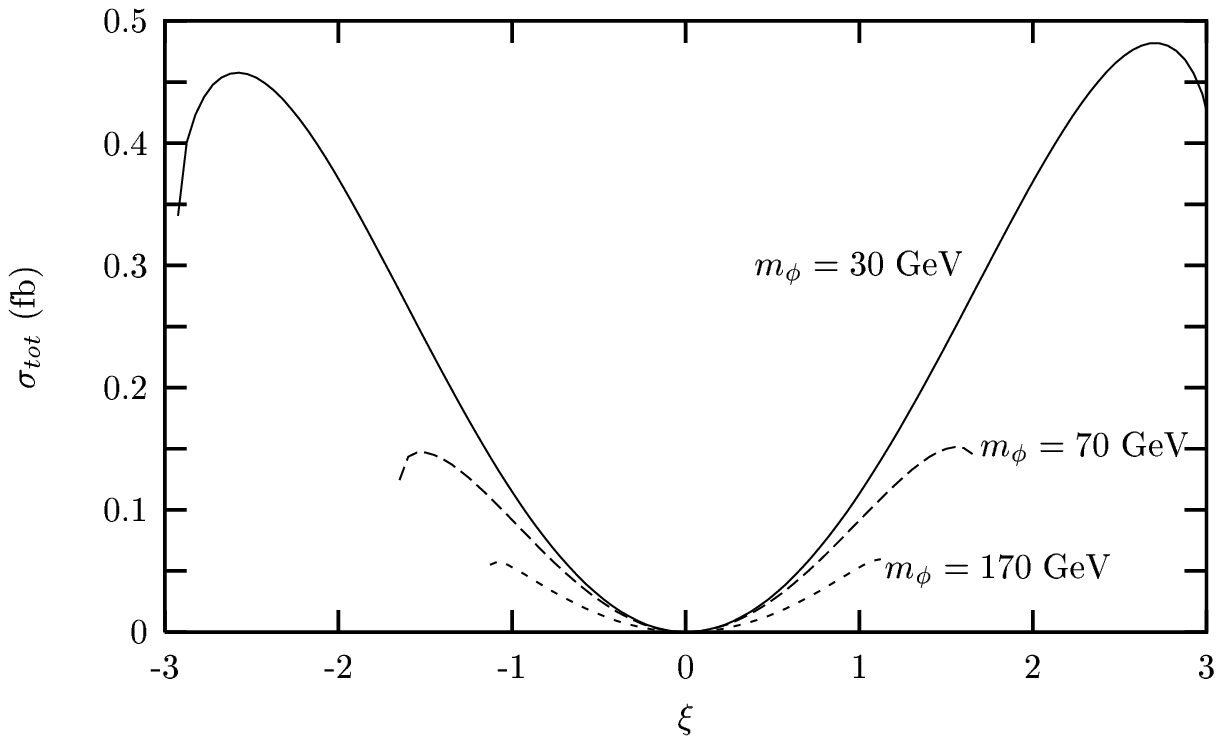}
\includegraphics[width=3in]{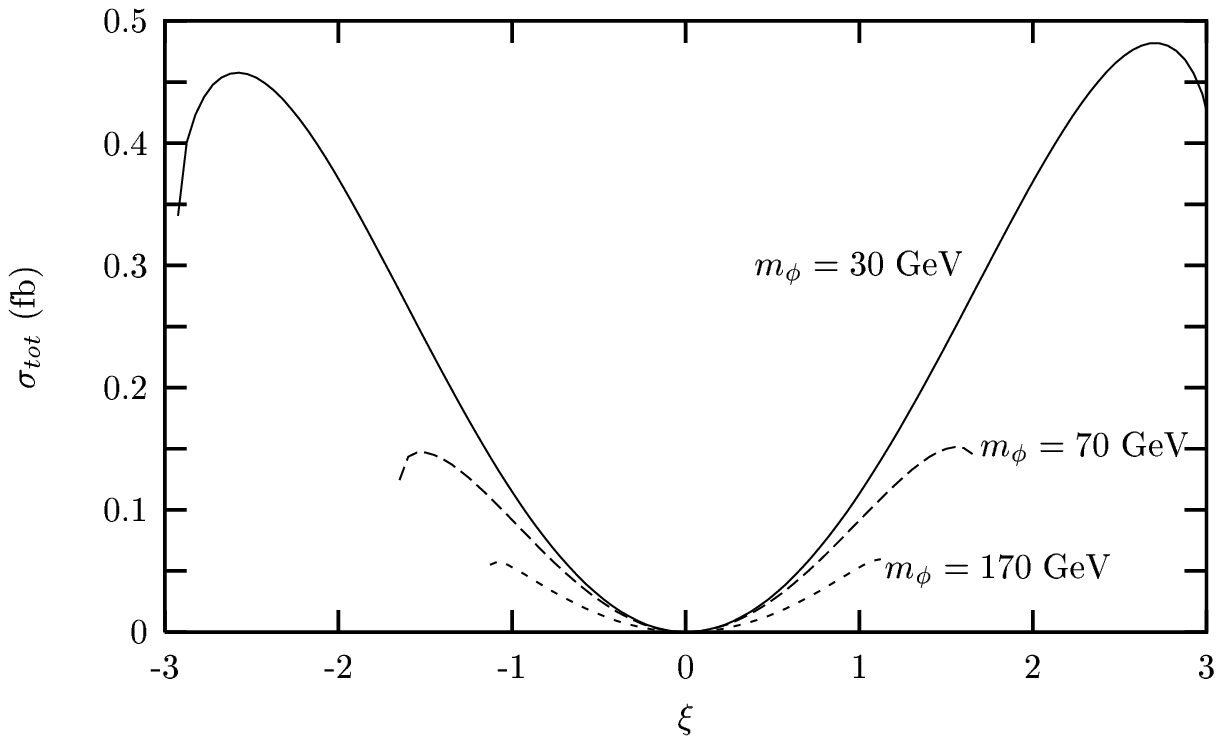}
\caption{\label{rs-9} The production of $e^+ e^- \to
G^{(n)} \to h \phi$ vs $\xi$, and (b) the constrained parameter
space in $(m_\phi, \xi)$. From Ref.~\protect\cite{RS-8}.}
\end{figure}

\begin{figure}[th!]
\centering
\includegraphics[width=4.5in]{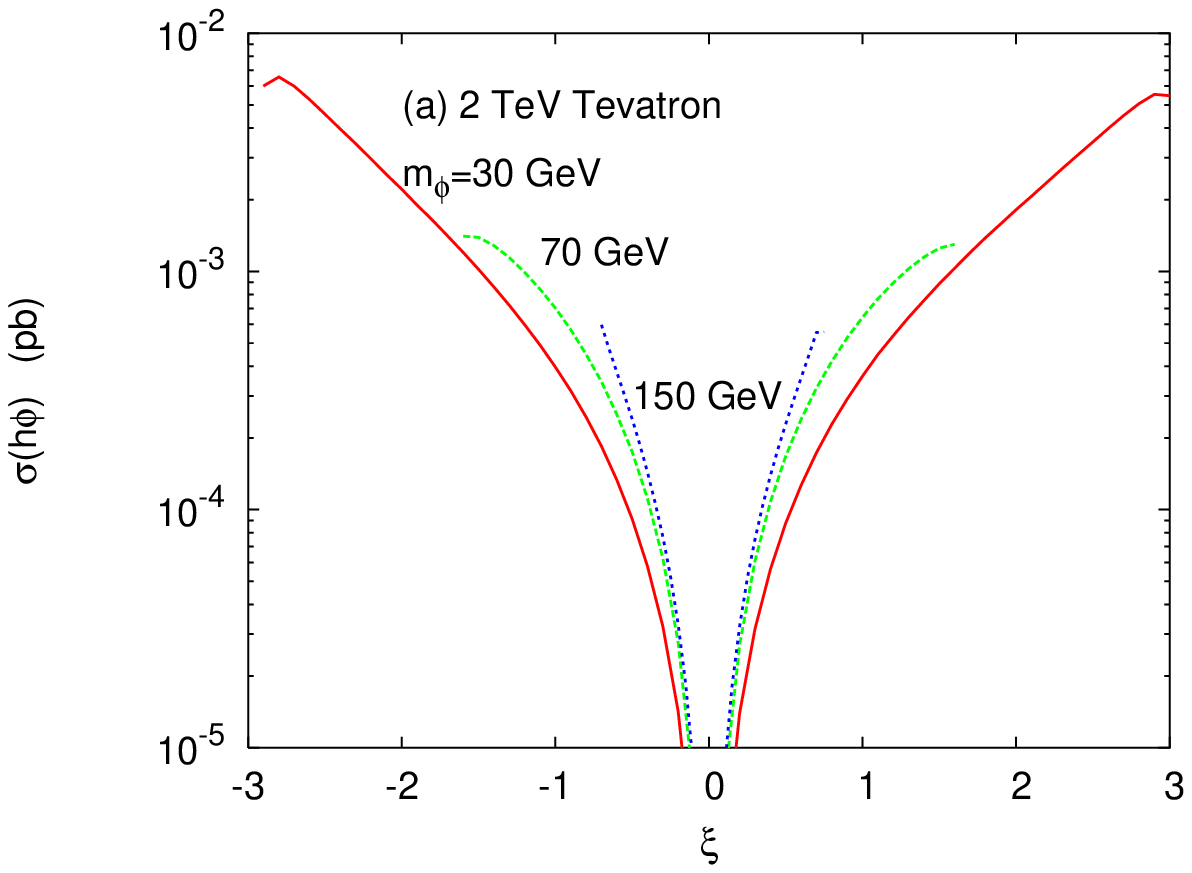}
\includegraphics[width=4.5in]{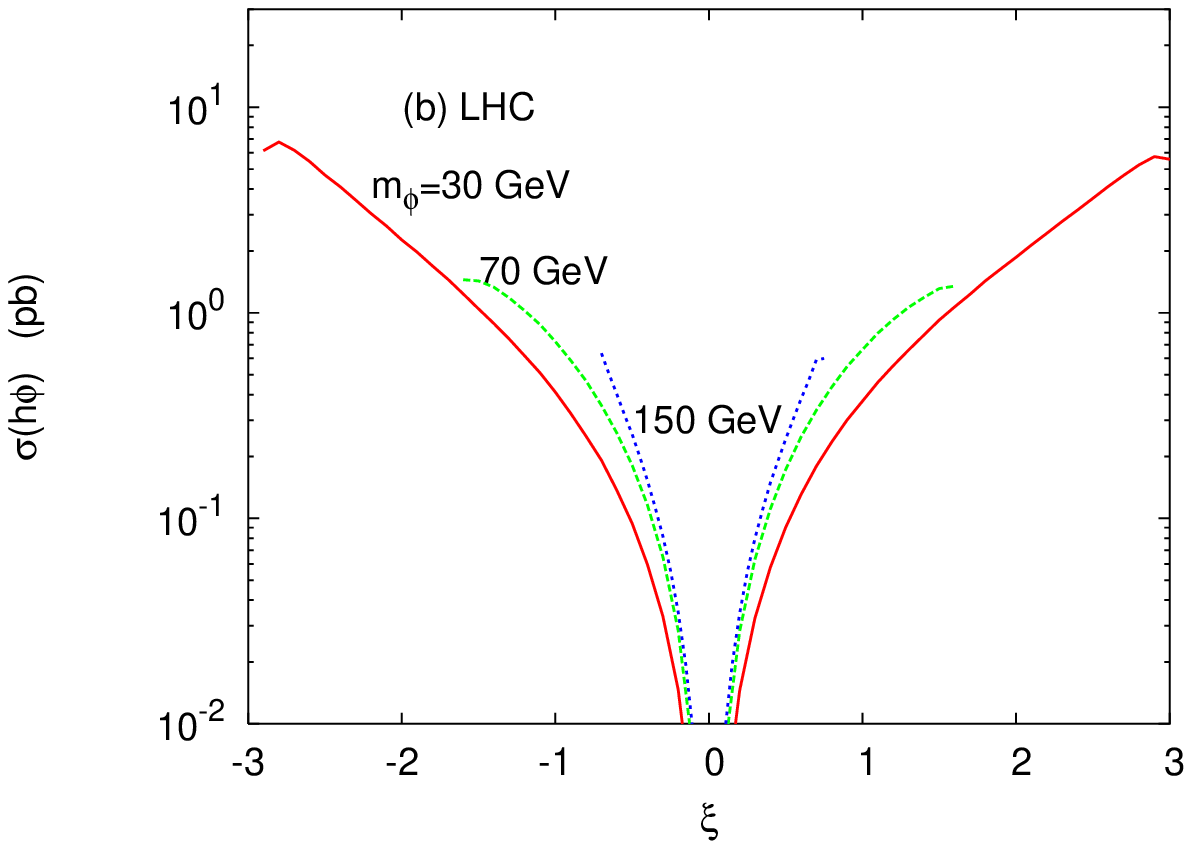}
\caption{ \label{rs-10} 
Cross sections for $h\phi$ production versus $\xi$ at the Tevatron 
and at the LHC.}
\end{figure}

Figure \ref{rs-8} shows the change in the branching ratios of the
Higgs boson and the radion vs $\xi$.  An interesting channel to
probe the mixing would be the observation of the triple couplings
mentioned above, e.g., in the process $e^+ e^- \to G^{(n)} \to h
\phi$ \cite{RS-8}.  Figure \ref{rs-9} shows the cross section of this process
vs $\xi$.  It  is obvious that the cross section goes to zero when
$\xi \to 0$.  The second part of the figure shows the region
in the parameter space that can be probed in the next linear
collider.  
We can also probe the mixing through the processes 
$gg,q\bar q \to G^{(n)} \to h \phi$ at hadron colliders \cite{RS-8}.  
We show the sensitivity of the cross section versus $\xi$ at the 
Tevatron and at the LHC in Fig.~\ref{rs-10}.
A partial list of other works in radion phenomenology are listed in 
Refs. \cite{RS-other}.

\section{TeV$^{-1}$-sized extra dimensions with gauge bosons}

This scenario was originally proposed by Antoniadis \cite{anto}.
Later, it was employed by Dienes et al. \cite{tev-1} to achieve
the early gauge coupling unification.   When the running scale
reaches the compactification scale of the extra dimensions, the
gauge couplings actually feel the strong presence of the KK states
of the gauge bosons.  The running of the gauge couplings will be
accelerated from a logarithmic running to a power-law running.
Thus, early unification is possible.

\begin{figure}[th!]
\centering
\includegraphics[width=3.1in]{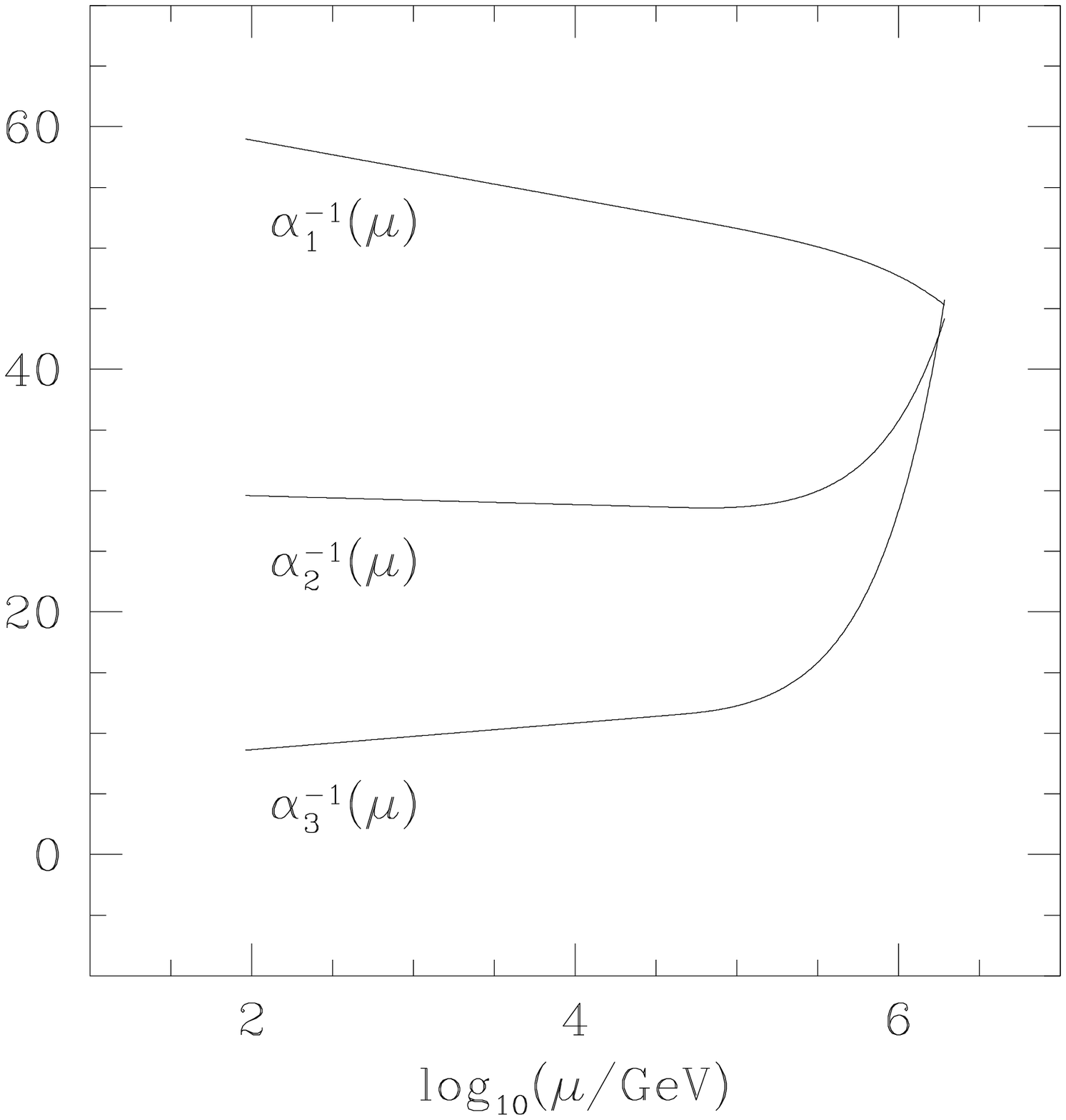}
\includegraphics[width=3.1in]{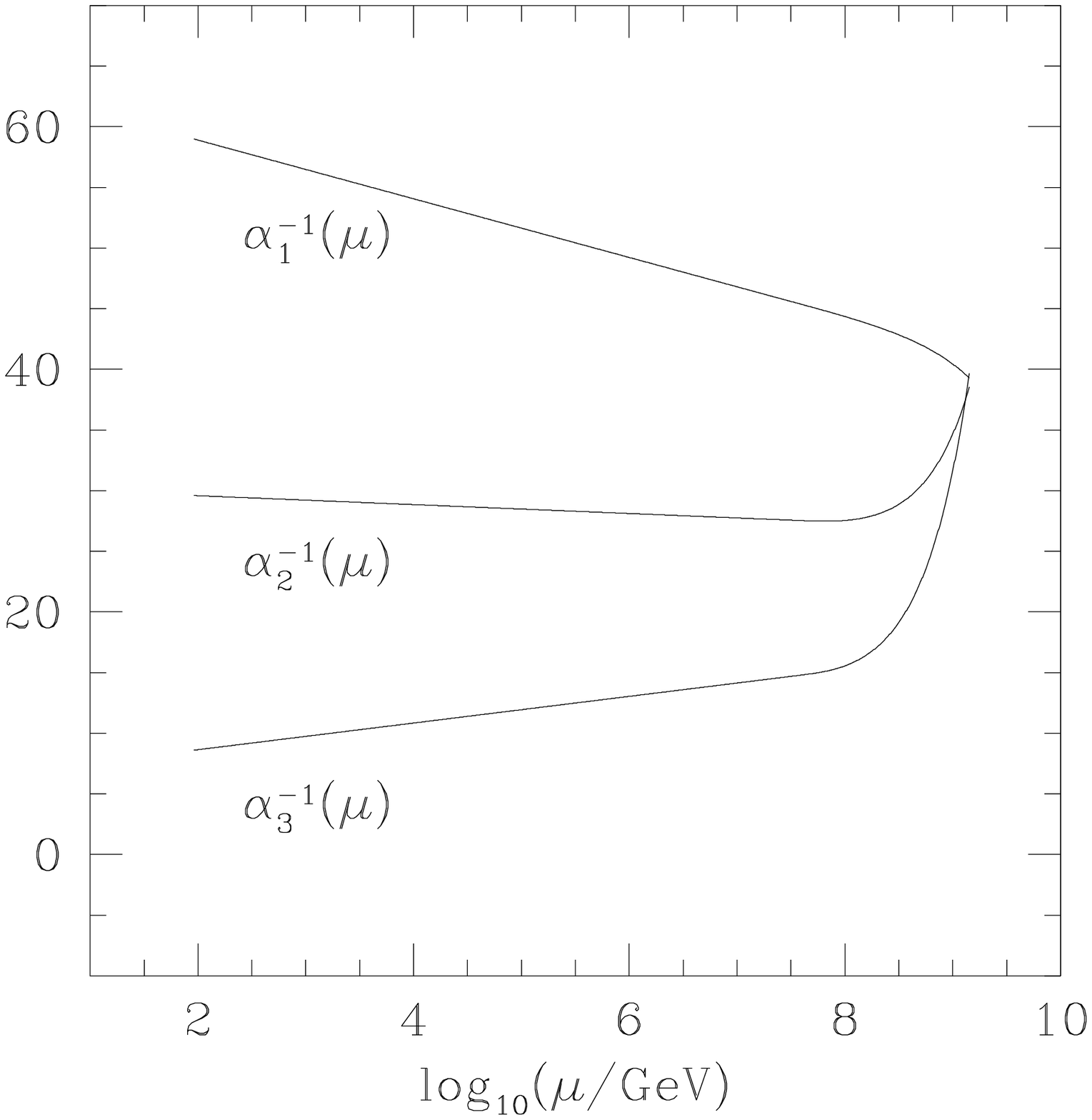}
\caption{  \label{tev-1} Early gauge coupling unification.
Here $\delta=1$, $\mu_0=10^5, 10^8$ GeV, respectively. This is taken from 
Ref.~\protect\cite{tev-1}. }
\end{figure}

 With the gauge bosons in the bulk, the KK states have masses
\[
m_n^2 = m_0^2 + \sum^\delta_i \frac{n_i^2}{R^2} \;.
\]
When the energy scale is above $\mu_0\equiv 1/R$, the KK states
contribute to physical processes, e.g., the running of the
couplings:
\begin{eqnarray}
\alpha_i^{-1} (\Lambda) &=& \alpha_i^{-1} (M_Z) - \frac{b_i}{2\pi}
\ln \frac{\Lambda}{M_Z} + \frac{\tilde{b_i}}{2\pi} \ln
\frac{\Lambda}{\mu_0} \nonumber \\
&& - \frac{\tilde{b_i} X_\delta}{2\pi \delta}
\left[ \left( \frac{\Lambda}{\mu_0} \right )^\delta - 1 \right ]
\end{eqnarray}
where
\begin{eqnarray}
(b_1,b_2,b_3) &=& (33/5,1,-3);\; (\tilde b_1, \tilde b_2, \tilde
b_3)= (3/5, -3, -6);\; \nonumber \\
&&   X_\delta = \frac{2\pi^{\delta/2}}
  {\delta \Gamma(\delta/2) } \;.
 \end{eqnarray}
Examples of early gauge coupling unification are shown in Fig. \ref{tev-1}.

Phenomenology of KK gauge bosons has been considered in a 5D model
with the extra dimension compactified on $S^1/Z_2$ \cite{tev-2}. 
The 5-D Lagrangian is given by
\begin{equation}
{\cal L}_5 = - \frac{1}{4g_5} F^2_{MN} + \left| D_M \phi_1 \right
|^2
 +\left( i \bar\psi \sigma^\mu D_\mu \psi + \left|D_\mu \phi_2 \right |^2
\right )\delta(x^5)
\end{equation}
 Compactifying the fifth dimension with
 \[
\Phi(x^\mu, x^5) = \sum_{n=0}^{\infty} \cos \left(\frac{n x^5}{R}
\right) \Phi^{(n)}(x^\mu)
\]
where $\Phi$ represents the gauge fields. The resulting 4-D
Lagrangian becomes
\begin{eqnarray}
{\cal L}^{\rm CC} &=& \frac{g^2 v^2}{8} \biggr [ W_1^2 +
\cos^2\beta
\sum_{n=1}^{\infty}( W_1^{(n)})^2 + 2\sqrt{2} \sin^2\beta W_1 
 \sum_{n=1}^{\infty} W_1^{(n)} + 2 \sin^2\beta
\left( \sum_{n=1}^{\infty} W_1^{(n)} \right )^2 \biggr ] \nonumber \\
&+& \frac{1}{2} \sum_{n=1}^{\infty} n^2 M_c^2 (W_1^{(n)})^2
 - {g} (W_1^\mu + \sqrt{2} \sum_{n=1}^{\infty} W_1^{(n)\mu} ) J_\mu^1
 +(1 \to 2)  \nonumber \\
{\cal L}^{\rm NC} &=& \frac{{g}v^2}{8 {c}_\theta^2} \biggr [ Z^2
+ \cos^2\beta \sum_{n=1}^{\infty}( Z^{(n)})^2 + 2\sqrt{2} \sin^2\beta Z 
  \sum_{n=1}^{\infty} Z^{(n)} + 2 \sin^2\beta \left
(\sum_{n=1}^{\infty}
 Z^{(n)} \right )^2 \nonumber \\
&+& \frac{1}{2} \sum_{n=1}^{\infty} n^2 M_c^2 \biggr[ (Z^{(n)} )^2
+
  (A^{(n)})^2 \biggr] \nonumber \\
&-& \frac{{e}}{ {s}_\theta {c}_\theta } \left (
 Z^\mu + \sqrt{2} \sum_{n=1}^{\infty} Z^{(n)\mu} \right ) J_\mu^Z
 - {e} \left (
 A^\mu + \sqrt{2} \sum_{n=1}^{\infty} A^{(n)\mu} \right ) J_\mu^{\rm em}
\nonumber
\end{eqnarray}
Here we explicitly write down the charged-current (CC) and neutral
current (NC) Lagrangians.

There are two types of phenomenology associated with the KK states
of the gauge bosons.  First, there will be mixings with the SM $W$
and $Z$ bosons \cite{tev-3}, because the KK states just have the same quantum
number as the SM gauge bosons. All the weak eigenstates mix to
form mass eigenstates, e.g., $Z^{(0)}$ mixes with all
$Z^{(n)}\;(n=1-\infty)$ through a series of mixing angles; similar
to $Z-Z'$ mixing. The lightest one is the $Z$ observed at LEP. The
couplings will be modified through the mixing angles. Thus, the
constraints from precision measurements place limits on $M_c$.
For example, in the presence of mixing, the Fermi constant and $Z$
decay partial widths are modified by
\begin{eqnarray}
G_F &=& \frac{\sqrt{2} g^2}{8 M_W^2} ( 1+ c_\theta^2 X) (1-2
\sin^2 \beta c_\theta^2 X ) \nonumber \\
\Gamma(Z\to f\bar f) &=& \frac{N_c M_Z}{12\pi}
\frac{e^2}{s_\theta^2 c_\theta^2}
 \, ( 1- 2 \sin^2\beta X) \,( g_v^2 + g_a^2) 
\nonumber \\
X &=& \frac{ \pi^2 M_Z^2}{3 M_c^2} \nonumber \;.
\end{eqnarray}
There have been a number of works using the electroweak precision
measurements \cite{tev-3} to place the constraint on
\[
M_c \agt 3.6\; {\rm TeV} \;.
\]

\begin{figure}[th!]
\centering
\includegraphics[angle=270,width=4.5in,clip]{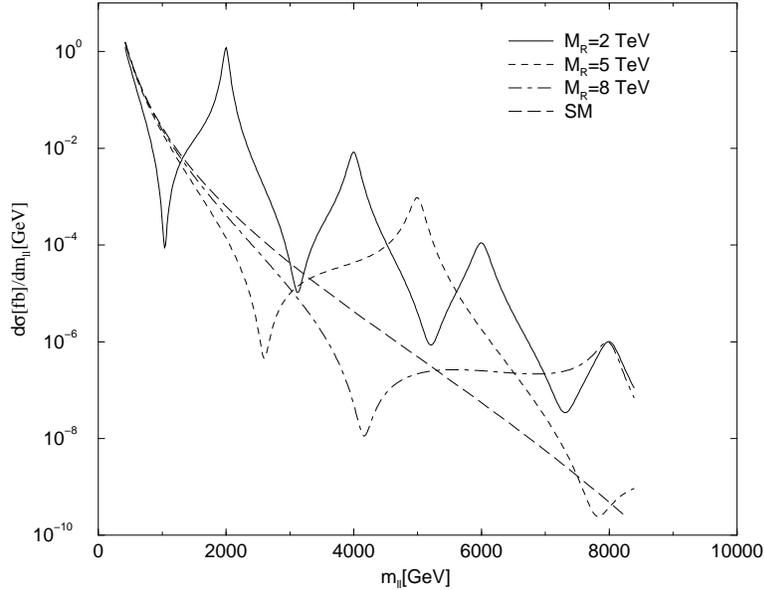}
\caption{ \label{tev-2} The KK resonances of photon and $Z$ 
in the Drell-Yan process. This is taken from Ref.~\protect\cite{tev-4}.}
\end{figure}

On the other hand, if the energy scale is higher than the
compactification scale $M_c$, resonances can be observed in
experiments, e.g., in the Drell-Yan production \cite{tev-4}:
see Fig. \ref{tev-2}.

If the energy scale is smaller than $M_c$, we should also expect
some virtual effects due to the KK states \cite{tev-5}.  Therefore, we can use
the existing high energy data to constrain the compactification
scale.
 In the approximation
\[
M^2_c \gg \hat s, |\hat t|, |\hat u| \;,
\]
the reduced amplitudes in the neutral-current scattering in $eq
\to eq$ can be obtained as \cite{tev-5}
\begin{eqnarray}
M^{eq}_{\alpha\beta}(s) &=& e^2 \Biggr \{ \frac{Q_e Q_q}{s} +
\frac{g_\alpha^e g_\beta^q}{\sin^2\theta_{\rm w} \cos^2
\theta_{\rm w} } \;
\frac{1}{s - M_Z^2 } \nonumber \\
&-& \left( Q_e Q_q +  \frac{g_\alpha^e g_\beta^q}
     {\sin^2\theta_{\rm w} \cos^2 \theta_{\rm w} } \right ) \;
 \frac{\pi^2}{3 M_c^2 } \; \Biggr \} \; \nonumber
\end{eqnarray}
where the first two terms are due to the photon and $Z$ exchanges
while the last term is due to the combined effect of the KK
photons and KK $Z$ bosons.  The effects can show up in the
interference terms and the pure KK term.

Cheung and Landsberg \cite{tev-5} used the following data sets in the analysis:
(i)  Drell-yan production at Tevatron, (ii) HERA NC and CC DIS,
(iii) LEPII hadronic, leptonic cross section, angular
distributions, (iv) dijet cross section and angular distribution,
and (v) $t\bar t$ production.  The limits obtained are shown in
Table \ref{table3}.  The overall limit is $M_c > 6.8$ TeV,
significantly improved from the electroweak precision data. We can
also estimate the sensitivity reach at the Run II of the Tevatron
and at the LHC \cite{tev-5} using the Drell-Yan process, 
shown in Table \ref{table4}.  A work by Bal\'{a}zs and Laforge \cite{balazs}
showed that using the dijet production the LHC can probe $M_c \sim 5-10$
TeV.
There are searches by D\O using dielectromagnetic objects \cite{kajf}
and by H1 \cite{kajf} on this model.  The limits they obtained are
$M_C > 1.12$ TeV and 1.0 TeV, respectively.

\begin{table}[th!]
\centering
\caption{ \label{table3} Best-fit values of
$\eta=\pi^2/(3M_c^2)$ and the 95\% C.L. upper limits on $M_c$ for
individual data set and combinations.} \medskip
\begin{tabular}{|l|cc|}
\hline
      & $\eta$ (TeV$^{-2}$)  &   $M_C^{95}$ (TeV) \\
\hline \hline
LEP~2:                       & & \\
{} hadronic cross section, ang. dist., $R_{b,c}$
   & $-0.33 \err{0.13}{0.13}$ & 5.3 \\
{} $\mu,\tau$ cross section \& ang. dist.
     & $0.09\err{0.18}{0.18}$ &  2.8 \\
{} $ee$ cross section \& ang. dist.
     & $-0.62\err{0.20}{0.20}$ & 4.5\\
{} combined          & $-0.28\err{0.092}{0.092}$ &
6.6 \\
\hline
HERA:     & & \\
{} NC     & $-2.74\err{1.49}{1.51}$ &  1.4 \\
{} CC     & $-0.057\err{1.28}{1.31}$ & 1.2 \\
{} HERA combined & $-1.23\err{0.98}{0.99}$  &1.6 \\
\hline
TEVATRON:              & &  \\
{} Drell-yan           & $-0.87\err{1.12}{1.03}$ &  1.3 \\
{} Tevatron dijet      & $0.46\err{0.37}{0.58}$ &  1.8 \\
{} Tevatron top production & $-0.53\err{0.51}{0.49}$ &  0.60 \\
{} Tevatron combined   & $-0.38\err{0.52}{0.48}$ & 2.3 \\
\hline \hline
All combined & $-0.29\err{0.090}{0.090}$ &6.8 \\
\hline
\end{tabular}
\end{table}

\begin{table}[th!]
\centering
\caption{ \label{table4} Sensitivity reach in $M_c$ for Run
1, Run 2 of the Tevatron and at the LHC, using the dilepton
channel. }
\medskip
\begin{tabular}{|c|c|}
\hline
     & 95\% C.L. lower limit on $M_C$ (TeV) \\
\hline \hline
{Run 1 (120 pb$^{-1}$)}&  1.4  \\
\hline
{Run 2a (2 fb$^{-1}$)} &  2.9 \\
\hline
Run 2b (15 fb$^{-1}$)  &  4.2 \\
\hline
{LHC (14 TeV, 100 fb$^{-1}$, 3\% systematics)}  & 13.5 \\
\hline LHC (14 TeV, 100 fb$^{-1}$, 1\% systematics) &
15.5 \\
\hline
\end{tabular}
\end{table}

\section{Universal Extra Dimensions}

In all previous models, all or part of the SM particles are confined to a
brane while some are free to move in the extra dimensions.  
This kind of models is in general 
easier to build because we are familiar with the $3+1$ 
dimensions for a long time.  However, there are no good reasons why we should
confine the SM particles on a 3-brane.  It is therefore appropriate to 
consider the scenario that all particles are free to move in all dimensions, 
dubbed universal extra dimensions \cite{ue-1}.  
Consider the case with only one extra dimension.
The momentum conservation in the fifth dimension, after compactification, 
becomes conservation in KK numbers (or called KK momentum).  
There may be some boundary terms arising
from the fixed points that break the conservation of KK numbers into a $Z_2$ 
parity, called KK parity.  Odd parities are assigned to 
the Kaluza-Klein states with an odd KK number.  Note that this breaking 
strength is at a few \% to about 10\% level compared to the SM coupling
strength.

\begin{figure}[th!]
\centering
\includegraphics[width=5in]{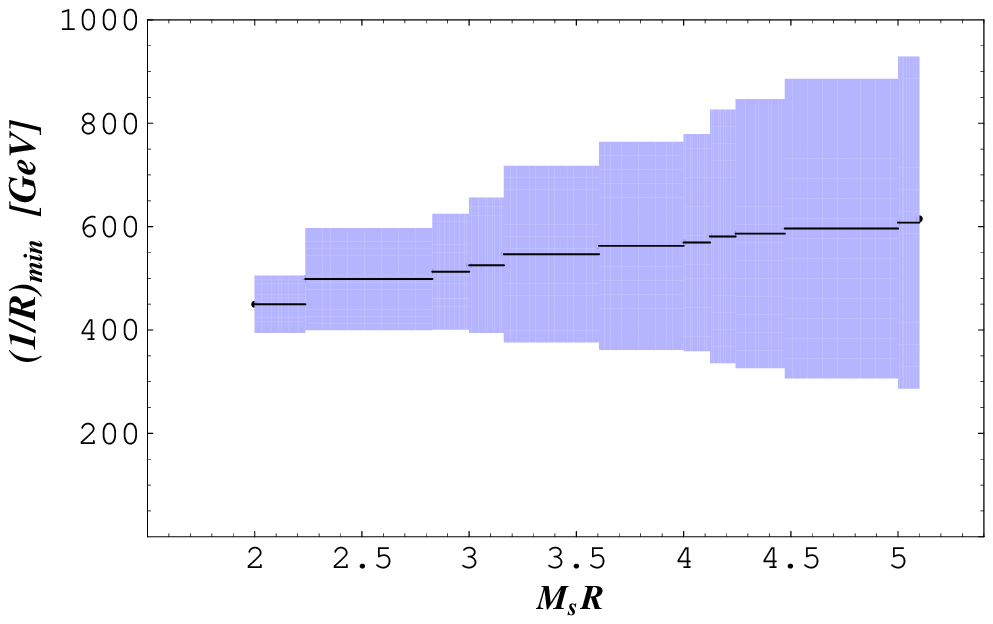}
\caption{\label{kk-constraint}
The present constraint on the universal extra dimension scenario for 
$\delta=2$ and $M_s$ is the upper cutoff.  This is taken from 
Ref.~\protect\cite{ue-1}.}
\end{figure}

Because of the KK number conservation (the size of KK number violating
couplings are much smaller) each interaction vertex involving KK states must 
consist of pairs of KK states of the same KK number.  
Therefore, in all processes
the KK states must exist in pairs including internal propagators.  This is
in contrast to all other scenarios like ADD, bulk gauge bosons, ...  Thus, 
the present limit on the universal extra dimension scenario is rather weak,
as weak as $1/R \agt 300$ GeV for one extra dimension \cite{ue-1} from 
precision data.  For the case of two extra dimensions the limit depends 
logarithmically on the cutoff scale $M_s$.  Roughly, the limit is around 
$400-800$ GeV, shown in Fig. \ref{kk-constraint}.

\begin{figure}[th!]
\centering
\includegraphics[width=3.1in]{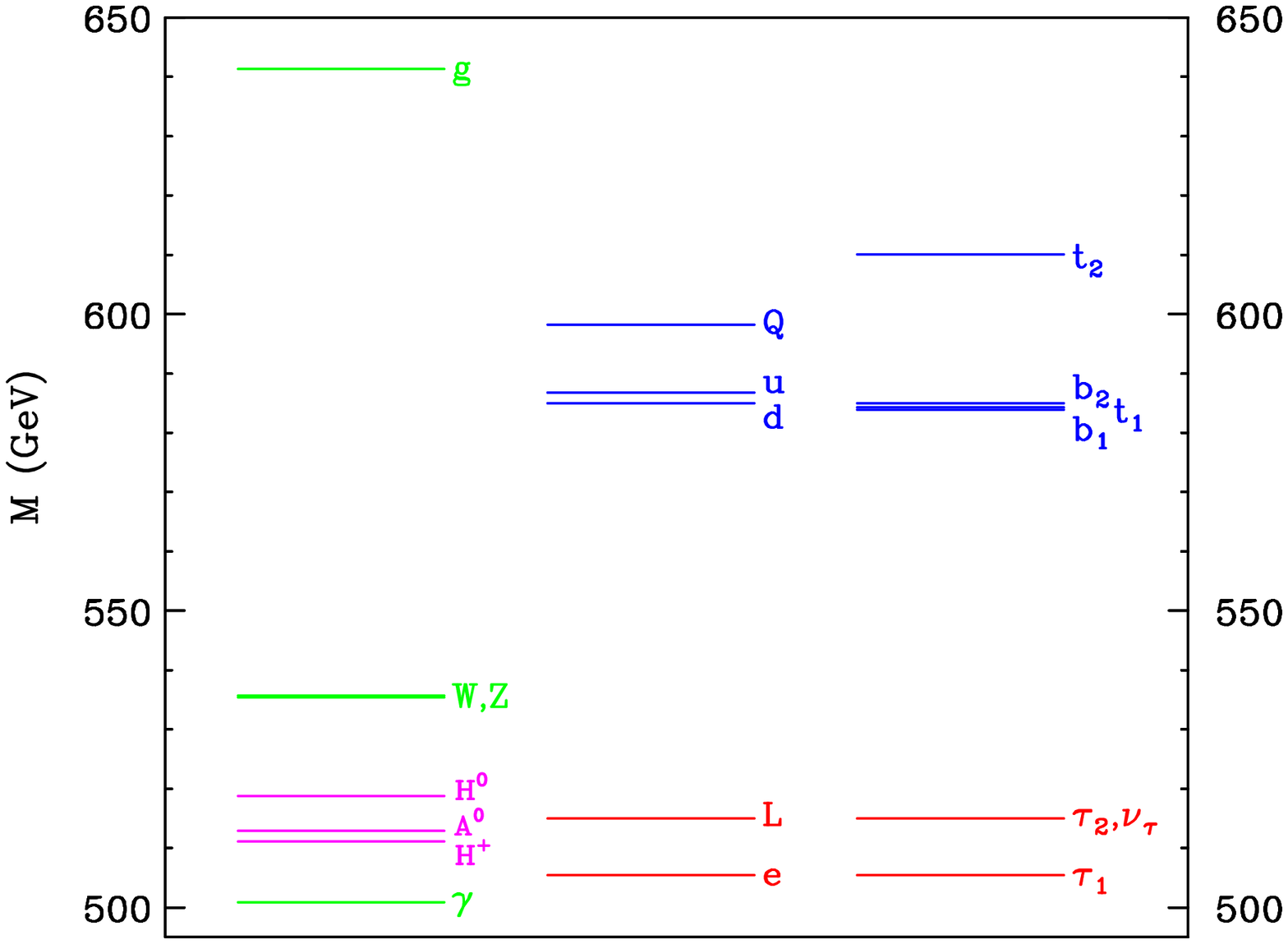}
\includegraphics[width=3.1in]{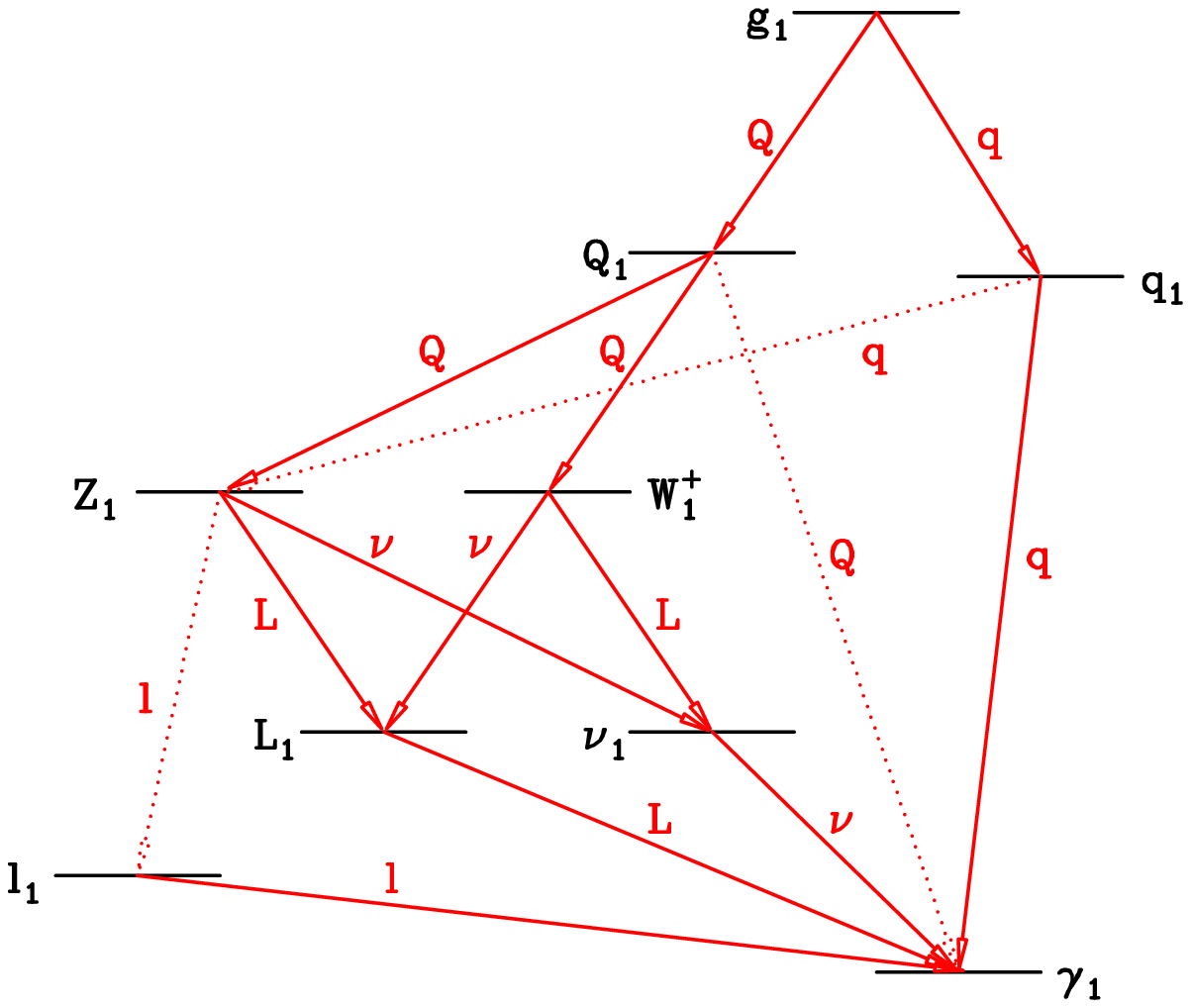}
\caption{ \label{kk-spectrum}
The mass spectrum and the possible decay chains of the first set of KK states
after taking into account the radiative corrections to the masses. 
These are taken from Ref.~\protect\cite{ue-3}.}
\end{figure}

The mass of the $n$th KK states is roughly
$n/R$, where $R$ is the compactified radius.  If there were no mass splitting
among the KK states of the same $n$, the phenomenology would be quite boring
that each $n=1$ KK state would be stable.  However, the radiative corrections
and boundary terms arised lift the mass degeneracy of the states of the same
$n$ \cite{ue-2}.  
The first KK state of the hypercharge gauge boson is the lightest KK
particle and it would be stable in collider experiments, and perhaps
stable over cosmological time scale because of the KK parity.  
Figure \ref{kk-spectrum} shows the spectrum and the possible decay chains 
of the first set of KK states
after taking into account the radiative corrections \cite{ue-3}.

\begin{figure}[th!]
\centering
\includegraphics[width=4.5in]{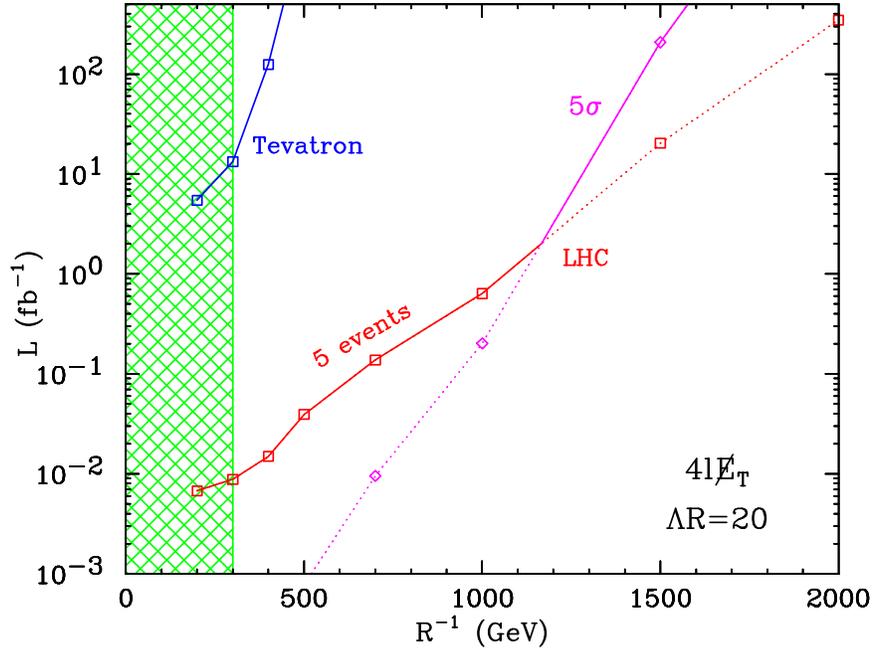}
\caption{ \label{kk-reach}
Sensitivity reach on the $1/R$ scale of the universal extra dimension
scenario at the RunII of the Tevatron and the LHC. This is taken from 
Ref.~\protect\cite{ue-3}.}
\end{figure}

Collider phenomenology is mainly the pair production of KK quarks and 
KK gluons \cite{ue-3}:
\begin{eqnarray}
q q' &\to& q^{(1)} q'^{(1)} \nonumber \\
q \bar q &\to& q^{(1)} \bar q^{(1)} \nonumber \\
gg &\to& g^{(1)} g^{(1)}\\ 
gg,q\bar q &\to& q^{'(1)} \bar q^{'(1)} \nonumber 
\end{eqnarray}
According to the decay chains shown in Fig. \ref{kk-spectrum}, each 
 $q^{(1)}$ decays into jets and $\gamma^{(1)}$ eventually, thus the
signature would be jets with missing energies.  The signal is very similar
to the weak-scale supersymmetry.  A more interesting decay chain 
would be that 
each $q^{(1)}$ decays into $W^{(1)}$ and $Z^{(1)}$, which in turn decay into 
leptons, thus giving rise to a signal of multi-leptons plus missing energies
\cite{ue-3}.
The sensitivity reach in the future collider experiments including Run II and
the LHC is shown in Fig. \ref{kk-reach}.  There is not much gain in the
Run II of the Tevatron, but the LHC with 100 fb$^{-1}$ integrated luminosity 
can probe up to $1/R \sim 1.5$ TeV.

\begin{figure}[th!]
\centering
\includegraphics[width=4.5in]{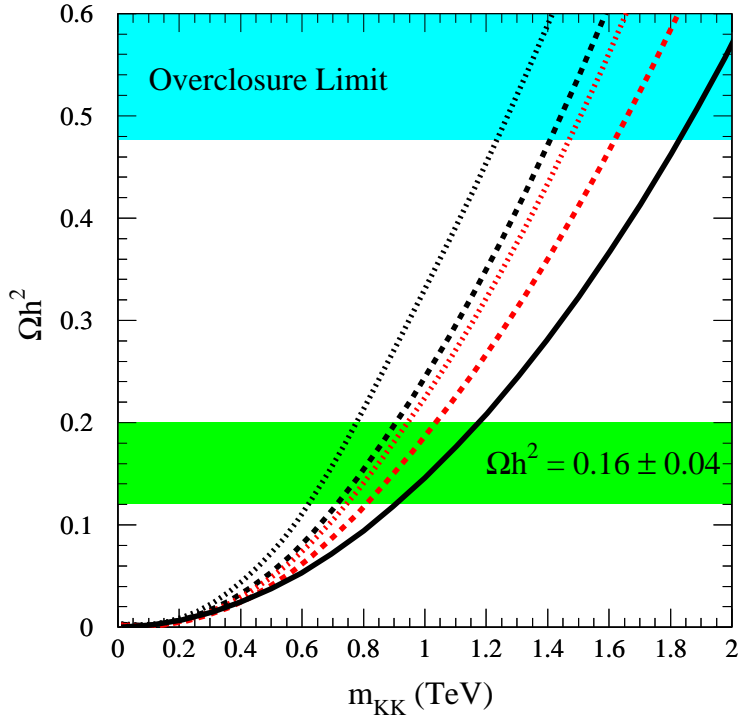}
\caption{ \label{kk-relic}
The relic density $\Omega h^2$ vs the mass of the lightest KK state $B^{(1)}$,
with or without the coannihilation effects. This is taken from
Ref.~\protect\cite{ue-4}.}
\end{figure}

\begin{figure}[th!]
\centering
\includegraphics[width=4.5in]{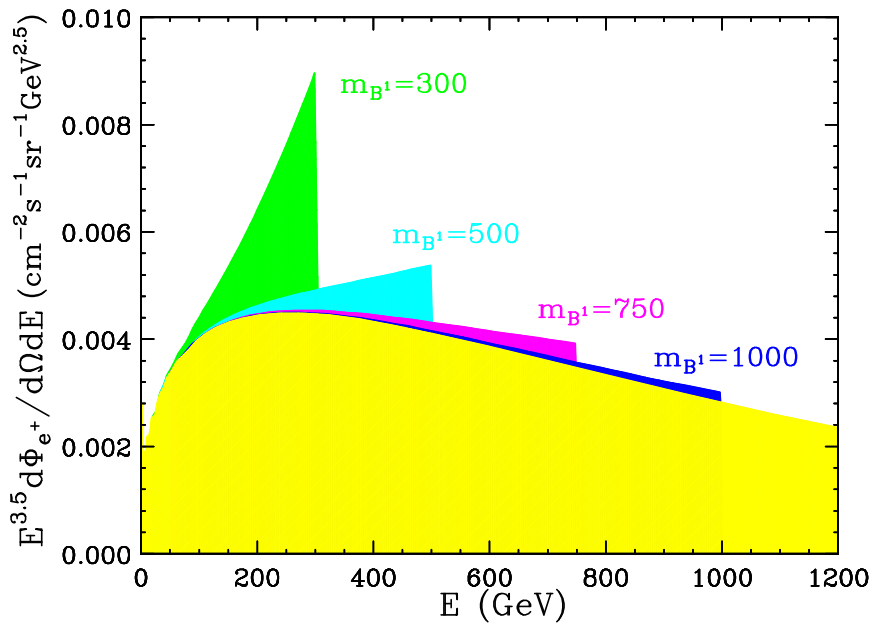}
\caption{\label{kk-dm}
The monoenergetic positron signal from the annihilation of the KK dark matter,
but broadened during propagation. This is taken from Ref.~\protect\cite{ue-5}.}
\end{figure}

Yet, the most interesting feature of the universal extra dimensions is 
the possible candidate of the cold dark matter -- the lightest KK state 
due to the KK parity.  Servant and Tait \cite{ue-4} 
calculated the relic density 
with or without the coannihilation channels in Fig. \ref{kk-relic}.
It is shown that for $B^{(1)}$ of order $800-1000$ GeV, it forms a major 
component of the cold dark matter.  A very striking signal of the KK
dark matter is the monoenergetic positron signal \cite{ue-5} 
from the annihilation of 
the cold dark matter, $B^{(1)} B^{(1)} \to e^+ e^-$, which can be detected
at, e.g., AMS experiment.  Figure \ref{kk-dm} shows the monoenergetic 
positron signal due to the annihilation of the KK dark matter.  Since the
$B^{(1)}$ pair annihilates into an electron-positron pair, the energy of the
positron is monoenergetic and equal to the mass of $B^{(1)}$. This is in 
sharp contrast to the positron signal of other cold dark matter candidates, 
e.g., the lightest neutralino.   However, the monoenergetic
spectrum would be broadened during the propagation to the Earth, but should
still be observable above the continuum background.

A partial list of other works on universal extra dimensions are listed
in Refs. \cite{ue-other}.

\section{An 5D SU(5) SUSY GUT Model}

This type of grand unified models in extra dimensions is based on orbifolding. 
By assignment of
different spatial parities (or boundary conditions) to various
components of a multiplet, the component fields can have very
different properties at the fixed points.  Thus, it is possible to
break a symmetry or to achieve the doublet-triplet splitting by the boundary
conditions.

\begin{figure}[th!]
\centering
\includegraphics[width=4.5in]{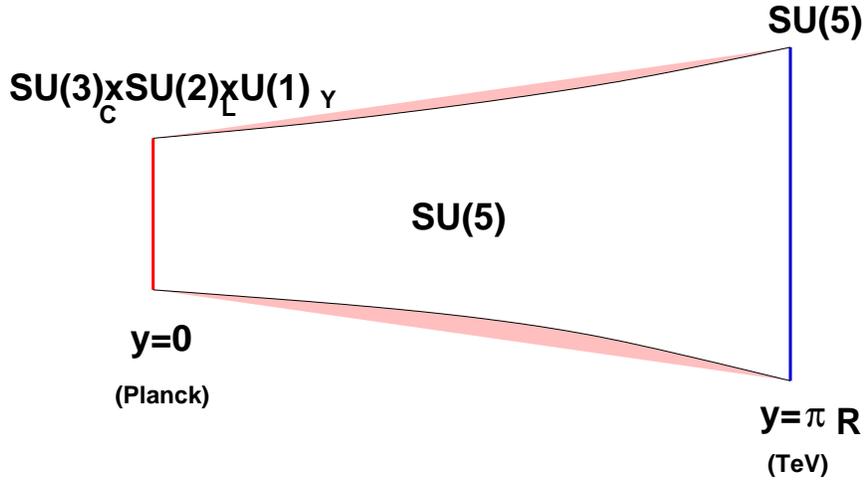}
\caption{\label{5d-su5}
A model of 5D SU(5) SUSY GUT on a slice of AdS space, due to
Goldberger, Nomura, and Smith \protect\cite{5d-1}.}
\end{figure}

In the model by Goldberger et al. \cite{5d-1}, they started from
the Randall-Sundrum scenario \cite{RS}: a slice of AdS space
with two branes (the Planck brane and the TeV brane), one at each end.
The hierarchy of scales is generated by the AdS warp factor $k$, which is
of order of the five-dimensional Planck scale $M_5$, such that the 4D
Planck scale is given by $M^2_{\rm Pl} \sim M_5^3/k$.  The fundamental
scale on the Planck brane is $M_{\rm Pl}$ while the fundamental scale
on the TeV brane is rescaled to TeV by the warp factor: $T\equiv k
e^{-\pi k R}$, where $R$ is the size of the extra dimension.
The setup is shown in Fig. \ref{5d-su5}.
The model is an 5D supersymmetric
SU(5) gauge theory compactified on the orbifold $S^1/Z_2$ in the AdS
space.  The boundary conditions break the SU(5) symmetry and provide
a natural mechanism for the Higgs doublet-triplet splitting and
suppress the proton decay \cite{5d-2}.  The Planck brane respects the SM gauge
symmetry while the TeV brane respects the SU(5) symmetry.  The matter
fermions reside on the Planck brane.  
By the boundary conditions the
wave-functions of the color-triplet Higgs fields are automatically zero
at the Planck brane, on which the matter fermions reside, while the
doublet Higgs fields are nonzero at the Planck brane and give Yukawa
couplings to the matter fermions.  Thus, the excessive proton decay
via the color-triplet Higgs fields is highly suppressed, and the
doublet-triplet splitting is therefore natural by the boundary conditions.
The mass of the color-triplet fields (and the XY gauge
bosons) is given by the warp factor and is of a TeV scale, the same as
the KK states of other fields in the setup.

\begin{figure}[th!]
\centering
\includegraphics[width=4.5in]{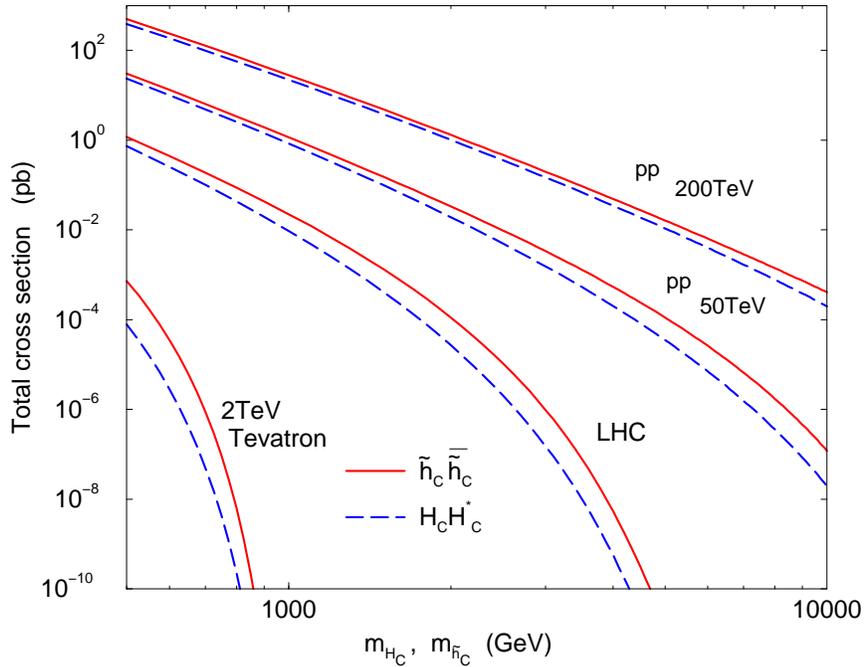}
\caption{ \label{cho-1}
Total production cross section for the colored Higgs bosons and Higgsinos
in $pp$ collisions. From Ref.~\protect\cite{5d-4}.}
\end{figure}

The striking signature of this type of GUT models is the TeV colored Higgs
bosons and colored Higgsinos \cite{5d-3,5d-4}, 
in contrast to the conventional proton decay signature.
Such TeV colored Higgs bosons and Higgsinos 
can be copiously produced at the upcoming LHC. 
Since these colored Higgs bosons and Higgsinos 
do not couple to matter fermions or
weak gauge bosons, they couple only to the gluons.  The interactions are
described by
\begin{eqnarray}
{\cal L} &=& -i g_s H_C^* {\stackrel{\leftrightarrow}{\partial}}_\mu 
H_C T^a A^{a\mu} 
     + g_s^2 T^a T^b H_C^* H_C A^a_\mu A^{b\mu} \nonumber \\
&& - g_s\, T^a A^a_\mu \,
       \overline{\widetilde{h}_C} \gamma^\mu   \widetilde{h}_C \, 
 -\sqrt{2} g_s \left(
H_C^* \, \overline{\widetilde{g}^a} \,T^a \, \widetilde{h}_C + 
\overline{\widetilde{h}_C} \,
 T^a \,\widetilde{g}^a \,  H_C \right) \;,
\label{cho-eq}
\end{eqnarray}
where $T^a$ is the generator of SU(3), and 
$A {\stackrel{\leftrightarrow}{\partial}}_\mu B 
\equiv A(\partial_\mu B) - (\partial_\mu A)B$.
Production is via the $q\bar q$ and $gg$ fusion.  The cross section 
formulas can be found in Ref. \cite{5d-3,5d-4}.  The total production cross
section is illustrated in Fig. \ref{cho-1}.

The detection of the colored Higgs boson depends on its decay modes.  From
Eq. (\ref{cho-eq}) it is clear that the colored Higgs boson must be coupled
pairwise to gluons and so itself cannot decay into gluons.  However, 
the colored Higgs boson will couple to its own supersymmetric partner, the
colored Higgsino, and the gluino or gravitino.  In general, we expect the
masses of the colored Higgs boson and the colored Higgsino to be of 
the same order.  Here we assume that the mass of the colored Higgs boson
is less than the sum of the masses of the colored Higgsino and the gluino
(or the gravitino), such that the colored Higgs boson is stable at
least within the detector.  (We can equally assume the reverse is true 
that the colored Higgsino is lighter than the colored Higgs bosons, 
then the following discussion will apply to the colored Higgsino.)

Once the colored Higgs bosons are produced, they will hadronize into 
massive stable particles, electrically either neutral or charged. For both
neutral or charged states, they will undergo very little hadronic energy
loss in the detector, because of the very small momentum transfer between the
Higgs boson and the detector material.  Therefore, the neutral state will
escape detection unnoticed.  The charged state will also undergo the ionization
energy loss though, through which it is detected.  The ionization energy loss
$dE/dx$ is very standard and can be found in Particle Data Book.  The 
$dE/dx$ almost has no explicit dependence on the mass of the particle. The
dependence comes in through the factor $\beta\gamma \equiv p/M$, in particular
for the range $0.1 < \beta\gamma <1$, $dE/dx$ is almost a linear function of
$\beta\gamma$ and has no dependence on the mass.  Therefore, by measuring
$dE/dx$ the $p/M$ can be deduced.  If the momentum $p$ is measured 
simultaneously, the mass $M$ of the particle can be estimated.

Experimentally, the massive stable charged particle will produce
a track in the central tracking and/or silicon vertex system, where $dE/dx$ 
and $p$ can be measured, provided that
$\beta \gamma$ is not too large ($\beta\gamma < 0.85$).
In the current search for stable charged particles, the CDF Collaboration
also required the particle to penetrate to the outer muon chamber.  This is 
possible if the initial $(\beta\gamma)_0 \agt 0.25- 0.5$.  Therefore,
we can call the signature the ``heavy muon''.  We have verified that
for a 1 TeV particle the requirement on $(\beta\gamma)_0$ is similar.

The event rate can be estimated including the following factors:
\begin{itemize}
\item The probability $P=1/2$
 for the colored Higgs boson to hadronize into
a charged state.
\item Require at least one colored Higgs boson
 to be in the detection range:
$0.25 < \beta\gamma < 0.85$ and $|\eta| < 2.5$.
\item An efficiency factor of 80\%
 for seeing a track in the central tracking
chamber.
\end{itemize}
Note that the $\beta\gamma \equiv p/M > 0.25$ cut means $p>250$ GeV for 
a 1 TeV particle, which makes it background free from 
$\mu^\pm, K^\pm, \pi^\pm$.  
The event rates are shown in Table \ref{rate}.  The RunII with a 20 fb$^{-1}$
is sensitive to a mass of about 400 GeV.  The LHC is sensitive up to about 
$1.5$ TeV \cite{5d-3}.  If including the feed-down from 
colored Higgsino production the sensitivity will improve to almost 2 TeV
\cite{5d-4}.

\begin{table}[th!]
\centering
\caption{ \label{rate}
Event rates for the production of colored Higgs bosons at the Tevatron, the
LHC and the future $pp$ colliders of 50 and 200 TeV.}
\medskip
\begin{tabular}{|c|cccc|}
\hline
$m_{H_C}$ (TeV) & Tevatron  & LHC   & VLHC 50 TeV & VLHC 200 TeV\\
   & (${\cal L}=20$ fb$^{-1}$) & (${\cal L}=100$ fb$^{-1}$) 
  & (${\cal L}=100$ fb$^{-1}$)  & (${\cal L}=100$ fb$^{-1}$) \\
\hline
$0.3$ & $160$  & $2.3\times 10^{5}$ & $3.1\times 10^{6}$ & $2.8\times 10^{7}$\\
$0.4$ & $14$
   & $5.5\times 10^{4}$ & $9.6\times 10^{5}$ & $9.7\times 10^{6}$\\
$0.5$ & $0.9$  & $1.7\times 10^{4}$ & $3.7\times 10^{5}$ & $4.3\times 10^{6}$\\
$0.8$ &  -     & $1200$             & $4.7\times 10^{4}$ & $7.1\times 10^{5}$\\
$1.0$ &  -     & $285$              & $1.6\times 10^{4}$ & $2.9\times 10^{5}$\\
$1.5$ &  -     & $15$
               & $2200$             & $5.6\times 10^{4}$\\
$2.0$ &  -     & $1.2$              & $470$              & $1.6\times 10^{4}$\\
$3.0$ &  -     & -                  & $43$               & $2700$\\
$4.0$ &  -     & -                  & $6.4$ & $690$\\
$6.0$ &  -     & -                  & -                  & $90$\\
$8.0$ &  -     & -                  & -                  & $19$\\
$9.0$ &  -     & -                  & -                  & $9.7$\\
\hline
\end{tabular}
\end{table}

\section{Conclusions}

There should not be any conclusions as this area is growing so fast that
interesting scenarios are popping up all the times.  We have to keep our eyes 
open for viable models.

So far, there have been extensive studies of sub-Planckian and 
trans-Planckian collider signatures for the large extra dimension model
(ADD model).  Experimentally, there are already some
limits around $M_D \sim 1 - 1.4$ TeV for the fundamental Planck scale.

The most striking feature of the Randall-Sundrum model is that it 
has a distinct unevenly spaced KK spectrum.  However, the first sign at
colliders is perhaps the radion or radion-Higgs mixing 
effects.

The TeV$^{-1}$-sized extra dimensions with gauge bosons 
will modify the gauge coupling running,
and affect the precision measurements,
and high energy scattering processes.  The current best limit is about
$M_c > 6.8$ TeV.  On the other hand, the scenario with every particle
in the extra dimensions (universal extra dimensions) has a very different
phenomenology.  The presence of KK number conservation
 renders that the KK particles
must be produced in pairs even in loop level.  Therefore, the present
limit is rather weak, of order of $300-800$ GeV from precision measurements.
The lightest KK state is stable over cosmological time scale and could be
a dark matter candidate if it has a mass around $800-1000$ GeV.

Finally, I have also mentioned an 5D SU(5) SUSY GUT model in AdS space,
which can have safe proton decay, a natural doublet-triplet splitting, and 
a TeV colored Higgs triplet.  The TeV colored Higgs boson becomes
an alternative signature for this kind of GUT, in contrast to proton decay.
The TeV colored Higgs boson can be copiously produced at hadronic colliders,
e.g. the LHC, and gives an interesting ``heavy muon''-like signature.  
The LHC is sensitive up to about 1.5 TeV.

I have benefitted a lot from other recent reviews 
\cite{review1,review0,review2,review3,review4,review5,kanti}
on these subjects.

\section{Acknowledgments}
I would like to thank D\O\ Collaboration, Feng, Shapere, Hewett,
Spiropulu, Davoudiasl, Rizzo, Dominici, Grzadkowski, Gunion,
Toharia, Dienes, Dudas, Gherghetta, Nath, Yamada, Yamaguchi, 
Appelquist, Cheng, Dobrescu, Matchev, Schmaltz, Servant, and Tait
for using their figures in this talk and report.  
I would also like to thank Wai-Yee Keung, Greg Landsberg, Chung-Hsien Chou,
C.S. Kim, Jeonghyeon Song, and Gi-Chol Cho for collaborations on various
parts of this talk and report.  
I also thank Kaoru Hagiwara for financial support.
Finally, I appologize to those whose works do not appear here due to 
space limitation.  The work was supported by NSC grants
92-2112-M-007-053- and 93-2112-M-007-025-.

\bibliographystyle{plain}

\end{document}